\documentclass[aps,prb,10pt,twocolumn,groupedaddress,floatfix,amsmath,amssymb,superscriptaddress]{revtex4-2}

\usepackage{amsmath}
\usepackage{amssymb}
\usepackage{braket}
\usepackage{graphicx}
\usepackage{float}
\usepackage{color}
\usepackage{siunitx}
\usepackage{gensymb}
\usepackage{bm}
\usepackage{xcolor}
\usepackage{footnote}
\usepackage{bbm}
\usepackage{tabularray}
\usepackage{array}

\newcommand{\mos}{MoS$_2$}

\newcommand{\vbg}{V_\mathrm{BG}}
\newcommand{\vtg}{V_\mathrm{TG}}

\begin{document}
	
	\author{Igor~Rozhansky}
	\affiliation{National Graphene Institute, University of Manchester, Manchester M13 9PL, United Kingdom} 
	\author{Michele~Masseroni}
	\affiliation{Laboratory for Solid State Physics, ETH Zürich, CH-8093 Zürich, Switzerland} 
	\author{Ricardo~Pisoni}
	\affiliation{Laboratory for Solid State Physics, ETH Zürich, CH-8093 Zürich, Switzerland} 
	\author{Suad~Alshammari}
	\affiliation{National Graphene Institute, University of Manchester, Manchester M13 9PL, United Kingdom} 
	\affiliation{Physics Department, Faculty of science, Northern Border University, Arar, Saudi Arabia}
	\author{Xue~Li}
	\affiliation{National Graphene Institute, University of Manchester, Manchester M13 9PL, United Kingdom} 
	\author{Thomas~Ihn}
	\affiliation{Laboratory for Solid State Physics, ETH Zürich, CH-8093 Zürich, Switzerland} 
	\author{Klaus~Ensslin}
	\affiliation{Laboratory for Solid State Physics, ETH Zürich, CH-8093 Zürich, Switzerland} 
	\author{James~McHugh}
	\affiliation{National Graphene Institute, University of Manchester, Manchester M13 9PL, United Kingdom} 
	\author{Vladimir~Fal'ko}
	\affiliation{National Graphene Institute, University of Manchester, Manchester M13 9PL, United Kingdom} 
	
	\email{igor.rozhanskiy@manchester.ac.uk}

	\title{Refined DFT recipe and renormalisation of band-edge parameters for electrons in monolayer MoS$_2$ informed by the measured spin-orbit splitting}
	
	\begin{abstract}
		Conduction band-edge spin-orbit splitting (SOS) in monolayer transition metal dichalcogenides determines a competition between bright and dark excitons and sets conditions for spintronics applications of these semiconductors. Here, we report the SOS measurement for electrons in monolayer MoS$_2$, found from the threshold density, $n_*$, for the upper spin-orbit-split band population, which exceeds by an order of magnitude the values expected from the conventional density functional theory (DFT). Theoretically, half of the observed value can be attributed to the exchange enhancement of SOS in a finite-density electron gas, but explaining the rest requires refining the DFT approach. As the conduction band SOS in MoS$_2$ is set by a delicate balance between the contribution of sulphur $p_x$ and $p_y$ orbitals and $d_{z^2}-d_{xz}$ and $d_{z^2}-d_{yz}$ mixing in molybdenum, we use a DFT+U+V framework for fine-tuning the orbital composition of the relevant band-edge states. An optimised choice of Hubbard U/V parameters produces close agreement with the experimentally observed conduction band SOS in MoS$_2$, simultaneously resulting in the valence-band SOS and the quasi-particle band gap which are closer to their values established in the earlier-published experiments. 
	\end{abstract}

	\maketitle
	\section{Introduction}
	Transition metal dichalcogenides (TMDs) are a promising group of two-dimensional semiconductors for a wide range of electronics~\cite{Radisavljevic2013MobilityEngineeringMoS2,Lembke2015SingleLayerMoS2Electronics,Dumcenco2015LargeAreaMoS2}, optoelectronics~\cite{Withers2015,Withers2015_2} and spintronics~\cite{Tang2019,PhysRevLett.119.137401,PERKINS2024205,Ahn2020} device applications~\cite{Manzeli2017,Regan2022,RevModPhys.90.021001,Ahn2020,OBrien2023,PERKINS2024205,Schaibley2016}. Among those, monolayer molybdenum disulphide (MoS$_2$) stands out because of its environmental stability, efficient electrostatic doping with the already-demonstrated high electron mobility, and the prospects of wafer-scale growth~\cite{yang_wafer_scale_mos2_2023,xia_12inch_mos2_2023,li_epitaxy_buffer_layer_2024,yu_eight_inch_mos2_2024,Kwon2024NatElecMoS2wafer,Kang2015NatureHighMobility,Aspiotis2023}. For that reason, an efficient modelling tool for the properties of charge carriers in MoS$_2$ is highly desirable, including both microscopic insight into their single-particle characteristics based on density functional theory (DFT) and mesoscale many-body analysis of electron-electron correlation effects. Although a massive volume of DFT coding has already been developed and realized in several 2D materials databases and reviews~\cite{Haastrup2018_C2DB,Zhou2019,Campi2023,Mounet2018,Choudhary2020_JARVIS,Shen2022_2DHT_Discovery,Lu2024_ML_2D_Review,Kormányos_2015}, a satisfactory fine-energy-scale agreement between the DFT-computed and experimentally-measured spin-orbit splitting (SOS) has yet to be achieved for monolayer MoS$_2$ conduction band (CB) electrons: its DFT-computed values are negligibly small (only a few meV) ~\cite{Kormanyos2013_MoS2_TW,PhysRevB.88.245436,Cheiwchanchamnangij2013_StrainSOC,RodriguesPela2024} as compared to the marks set by various experiments ($\ge$ 10 meV)~\cite{Marinov2017,PhysRevLett.121.247701,Lin2019_MoS2_ValleySusceptibility,doi:10.1021/acs.jpclett.3c02431,Robert2020}.
	
\begin{figure*}
		\centering		\includegraphics[width=0.9\textwidth]{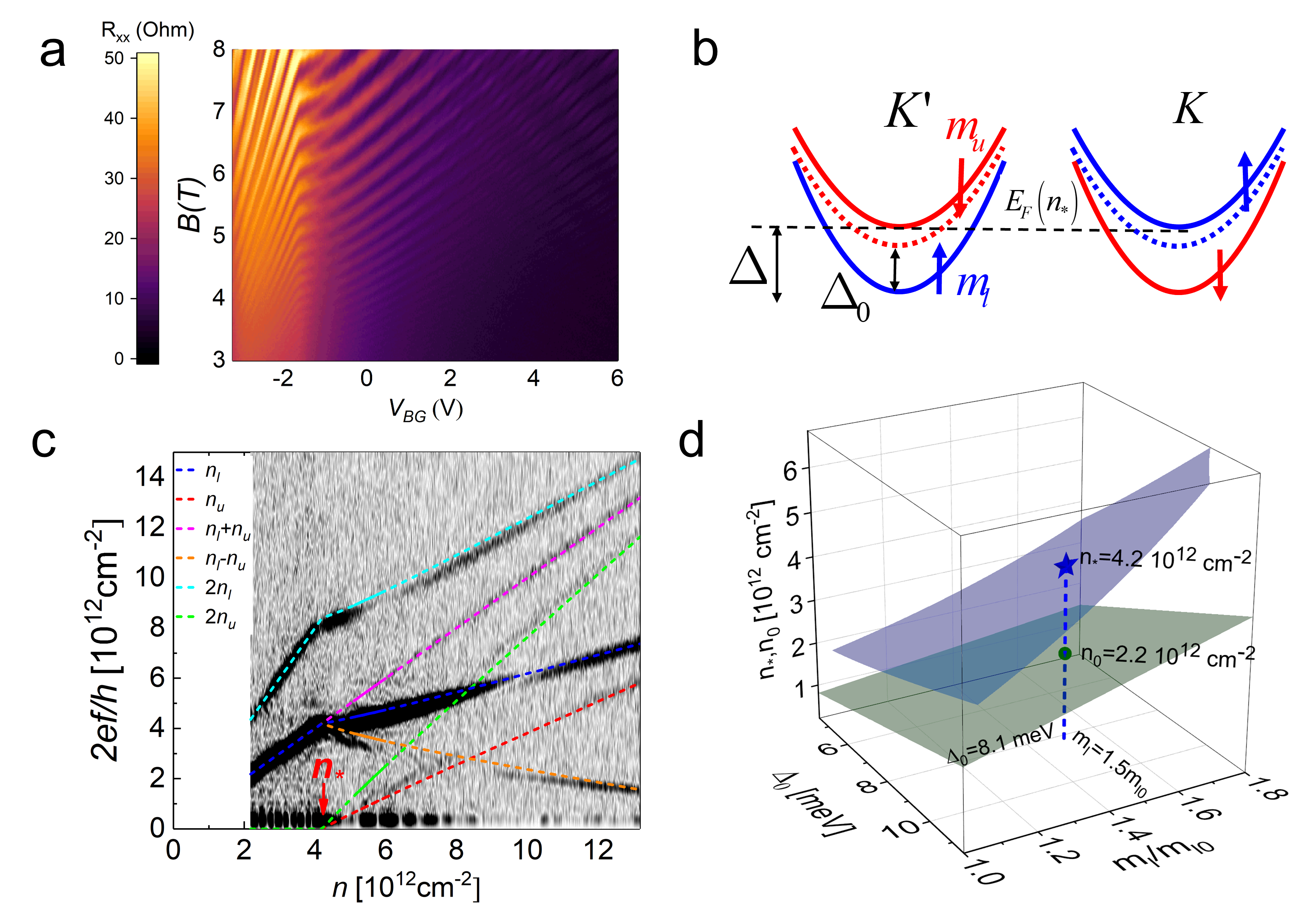}
		\caption{{\bf Comparison of experimental data and theoretical predictions for spin-split band fillings}.
			(a) Four-terminal resistance $R_{xx}$ as a function of the bottom gate voltage $V_{BG}$ and magnetic field $B$.
			(b) Band diagram indicating SOS conduction band of \mos~monolayer, effective masses $m_u$ and $m_l$ (red: spin down, blue: spin up) at the $\pm K$  valleys. The bare SOS $\Delta_0$ (dotted) increases to $\Delta$ (solid) by exchange interaction. The dashed horizontal line marks the Fermi level $E_F$ at the upper band filling threshold density, $n_*$. 
			(c) Density, $n$, dependence of the Fourier spectrum of Shubnikov-de Haas oscillations with the frequency $f*2e/h$ rescaled to represent electron sheet densities in the lower ($n_l$) and upper ($n_u$) spin-split bands, along with their combinations. The threshold density $n_*=4.2\times10^{12}$ cm$^{-2}$ marks onset of populating the upper spin-split band.
			At high-density end, $n_l$ and $n_u$ change almost linearly with $n$; from their slopes we estimate $m_u/m_l\approx1.3$ as the ratio between the upper and lower spin-split band masses. One may notice deviations from linear behaviour of $n_u$ near $n_*$, which is attributed to the electron-electron interaction as discussed in the Supplemental Information~\cite{supp} and with the calculated results shown by dashed lines.
			(d) 
			The  threshold density $n_*$ (blue surface) calculated from Eq.~(\ref{eqnstar}) as a function of the lower band effective mass $m_l$ (normalised by the DFT-computed value $m_{l0} \approx 0.43 m_e$) and bare SOS, $\Delta_0$, in comparison to threshold density $n_0$ evaluated without exchange renormalisation of SOS (green surface).
		}
		\label{fign1n2}
	\end{figure*}

	The purpose of the present study is to perform a detailed comparison of the experimentally-measured characteristics of spin-orbit-split electron bands in monolayer MoS$_2$ (Fig.~\ref{fign1n2}b) with the results of multi-scale modelling of those characteristics: the CB SOS, $\Delta$, and effective masses at the K-point band edges of MoS$_2$ Brillouin zone. Experimentally, as described in Section~\ref{secexper} these characteristics are determined by analysing the measured Shubnikov de Haas oscillations (SdHO, Fig.~\ref{fign1n2}a), enabling us to find the band mass of electrons and determine the threshold doping density, $n_*$, at which electrons filling reaches the bottom of the higher spin-orbit-split band, identified using the SdHO Fourier spectrum in Fig.~\ref{fign1n2}c. 
	
	The multi-scale modelling part of this work incorporates two components. First, in Section \ref{secee} we analyse many-body enhancement of a single-particle SOS, $\Delta_0$: it has recently been shown that 'bare' SOS is enhanced by exchange interaction between electrons in 2D semiconductors \cite{PhysRevB.110.L161404,Scharf_2019}, leading to the density dependence culminating at the threshold density, $n_*$. To mention, the threshold density $n_*$ is important for spintronics based on MoS$_2$-based FETs, as it separates the regimes of slow and fast spin and valley relaxation of charge carriers~\cite{PhysRevB.90.235429,PhysRevB.110.L161404}.  This enables us to determine $\Delta_0$ in monolayer MoS$_2$ from the experimentally measured $n_*$ value and, then, to compare it with the DFT results. Based on that comparison, in Section~\ref{secDFT}, we propose an extension of the commonly-used DFT approach to include both on-site ($U$) and inter-site ($V$ - between metal and chalcogen) Hubbard terms within DFT+$U$+$V$ in order to reproduce the extracted bare SOS value, simultaneously complying with the valence band SOS and the quasi-particle band gap, $E_g$.   
	\section{Magnetotransport data and S\lowercase{d}HO analysis}
	\label{secexper}
	The reported  experiments were conducted on a dual-gated multi-terminal \mos~device. The sample consists of a monolayer \mos~- encapsulated in hexagonal boron nitride (hBN), with pre-patterned metallic bottom contacts and graphite top and bottom gates. This sample is the same as the one presented in Ref.~\cite{PhysRevLett.121.247701}, and its fabrication details are provided in the Supplemental Information~\cite{supp}. The measured four-terminal resistance, $R_{xx}$ is shown in Fig.~\ref{fign1n2}a as a function of the bottom gate voltage, $\vbg$, and the perpendicular magnetic field, $B$. For this measurement, the top-gate voltage was set to $\vtg=\SI{8}{V}$ to ensure ohmic contact behaviour. The observed resistance exhibits clear SdHO that can be precisely described at low densities in terms of a single band of electrons with simple Kramers degeneracy. As reported earlier~\cite{PhysRevLett.121.247701},  temperature dependence of the amplitudes of these low-density SdHO is fitted well by effective mass $m_l=0.65 m_e$, attributed to the lower spin-split subband electrons, which is about 1.5 times bigger than typical values, $m_{l0}\approx 0.43m_e$, obtained in DFT calculations~\cite{Kormányos_2015,Kormanyos2013_MoS2_TW,PhysRevB.88.245436,Cheiwchanchamnangij2013_StrainSOC,RodriguesPela2024}. Note that this value is consistent with other published experimental data~\cite{Eknapakul2014}.
	
	At the higher densities, corresponding to the gate voltage $V_{BG}>V_*$, the SdHO pattern acquires a pronounced beating pattern and the SdHO fan plot changes its gradients (a 'kink'), indicating that, at $V_{BG}>V_*$, the total electron density is shared by two spin-split bands~\cite{PhysRevLett.121.247701}. As a result, from the value of $V_*$ in Fig.~\ref{fign1n2}a we determine the critical value, $n_*=4.2\times10^{12}~\text{cm}^{-2}$, of the electron sheet density at which the Fermi level in MoS$_2$ reaches the bottom of its upper spin-split band. The onset of the upper spin-split band occupation by electrons is even more obvious in the Fourier Transform of the $R_{xx}(B^{-1})$ data (performed for each $\vbg$ with the Fourier frequency rescaled into 'density units', as $2ef/h$), then plotted as a function of the total gate-controlled density, $n$, in Fig.~\ref{fign1n2}c. Note that this Fourier spectrum enables us to determine the carrier densities in both lower, $n_l$, and upper, $n_u$, spin-split bands, traced with blue and red dashed lines, respectively, and the identified density value, $n_*$, determined from Fig.~\ref{fign1n2}c raises an immediate issue, as the estimate for the upper spin-split band occupancy threshold using $\Delta_0 \approx 3$ meV and $m_{l0}\approx 0.43m_e$, as typically computed by DFT for undoped material, gives a threshold density $\Delta_0 m_{l0}/\pi\hbar^2\approx5\times 10^{11}\,\mathrm{cm^{-2}}$, which is about an order of magnitude below the experimentally measured value.  
	
	\section{Renormalization of bare band parameters by electron-electron interaction}
	\label{secee}

	A meaningful comparison between DFT-based band parameters and the experimentally determined threshold $n_*$ requires accounting for correlation-induced effects changing both SOS and the band mass~\cite{PhysRevB.110.L161404, Giuliani_Vignale_2005,Scharf_2019, PhysRevB.69.125334,PhysRevB.70.035104,PhysRevB.70.035111}, in particular, since the Wigner-Seitz radius is large throughout the density range covered by the data set in Fig.~\ref{fign1n2}c, up to $n_*$. On the one hand, as the experimental data in Section~\ref{secexper} offer the actual - that is, renormalised - band edge mass, we approximate the dispersion of the lower spin-split conduction band as parabolic, with the experimentally determined mass $m_l \approx 0.65\,m_e$, which sets the relation between doping density and the Fermi energy of electrons. The validity of the parabolic approximation is supported by the nearly constant SdHO period while filling the lower band. 
	On the other hand, we fully account for the many-body renormalisation of SOS and use it to determine the values of bare SOS, $\Delta_0$, from the experimentally identified threshold density, $n_* = 4.2 \times 10^{12} \, \text{cm}^{-2}$. To mention, that part of the calculation is based on the use of a bare band-edge mass, $m_{l0} \approx 0.43 m_e$, consistently found in various DFT calculations~\cite{Kormányos_2015,Kormanyos2013_MoS2_TW,PhysRevB.88.245436,Cheiwchanchamnangij2013_StrainSOC,RodriguesPela2024}.

	To realise the above-described strategy, we analyse the screened-exchange electron self-energy,   
	\begin{align}
		&\Sigma \left( {{\bf{k}},\omega } \right) = -\int {\frac{{{d^2}qd\Omega }}{{{{\left( {2\pi } \right)}^3}}}} \frac{{V\left( q \right)}G({\bf{k}} - {\bf{q}},\omega  - \Omega )}{{1 + V\left( q \right)\Pi \left( {q,\Omega } \right)}};
		\nonumber \\
		&V\left( {q} \right) = \frac{{2\pi {e^2}/\varepsilon }}{{q\left( {1 + {r_*}q} \right)}}
		;\quad r_* = \frac{{{\epsilon_\parallel } - \epsilon_{\bot }^{ - 1}{\varepsilon ^2}}}{{2\varepsilon }}d. \label{eqselfen}
	\end{align}	
	Here, we implement 2D dielectric screening via the Keldysh potential 
	(with screening length~\cite{GALIAUTDINOV20193167,PhysRevB.101.245432}, $r_*$, dependent on the film thickness, $d$) and metallic screening; $G({\bf k},\omega)=\left[\omega-(\hbar^2k^2/2m_{l0}-E_F)\right]^{-1}$ is the free valley-polarised electron Green's function in the lower SOS band; $\Pi({\bf q},\Omega)$ is the polarisation bubble describing metallic screening; and $\varepsilon=\sqrt{\varepsilon_\parallel\varepsilon_\bot}$ stands for the dielectric constant of the encapsulation environment (hBN) with in/out-of-plane dielectric constants, $\varepsilon_\parallel$/$\varepsilon_\bot$. 
	We use an effective dielectric constant for the hBN-encapsulated environment of $\varepsilon = 4.6$~\cite{Pierret_2022};
	For the in-plane and out-of-plane dielectric constant for MoS$_2$ monolayer we take the values~\cite{PhysRevB.92.205418,Menendez-Proupin2024} ${\epsilon _\parallel} = 15.46$, ${\epsilon_\bot} = 6.46$, $d = 5.48$ $ {\rm \AA}$, resulting in $r_* \varepsilon \approx 33$ ${\rm \AA}$.
	
\begin{figure}
		\centering   \includegraphics[width=0.49\textwidth]{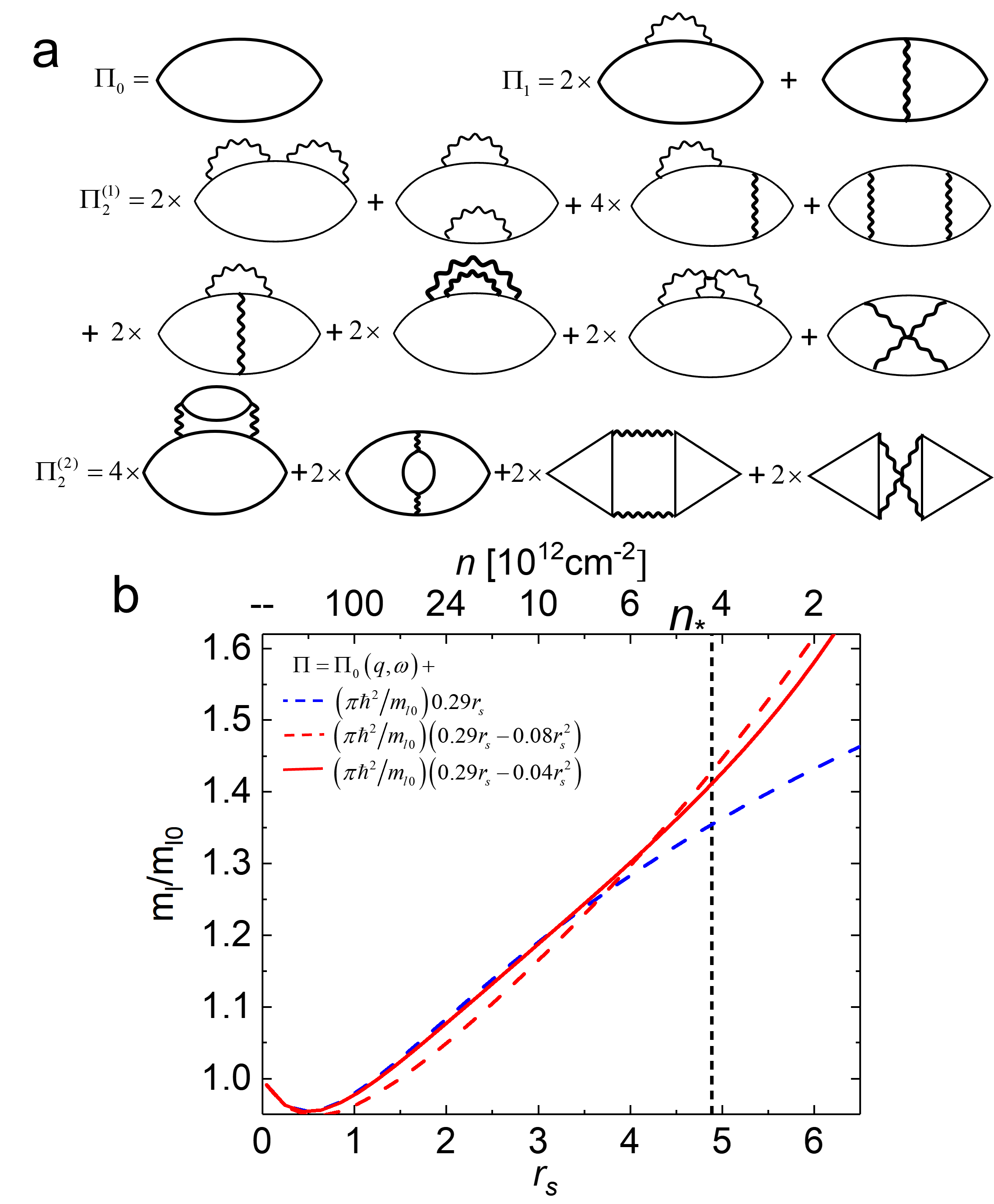}
		\caption{{(a) Set of diagrams contributing to the polarisability.}
			(b) 
			Calculated mass renormalisation. 
			The dashed blue and red lines show the mass renormalisation using a static first- and second-order correction to the polarisability, respectively. Red solid line is calculated with the median $\Pi(q,\omega)=\Pi_0(q,\omega)+(0.29r_s-0.04r_s^2)\frac{\pi\hbar^2}{m_{l0}}$; 
			the top and bottom axis show the electron density and corresponding Wigner-Seitz radius $r_s$, vertical dashed line corresponds to $n_*=4.2\times 10^{12} \text{ cm}^{-2}, \,r_s(n_*)=4.9$. }
		\label{figalldiags}
	\end{figure}

	At the threshold density $n_*$, Fermi level in the occupied lower band reaches the bottom of the upper spin-split band.
	Using the experimentally determined lower band mass, $m_l$, this condition reads as,
	\begin{equation}   
		\frac{{\pi {\hbar ^2}{n_*}}}{{{m_l}}} = \Delta \left( {{n_*}} \right) \equiv {\Delta _0} + \Sigma \left( 0,0 \right).
		\label{eqnstar}      
	\end{equation}
	Here, density-dependent self-energy correction is computed with statically screened interaction, implemented with $\Pi(q,0)$ at T=0. While for non-interacting 2D electrons $\Pi(q,0)$ is determined only by the bare mass, $m_{l0}$, electron-electron correlations introduce~\cite{Giuliani_Vignale_2005} density dependence of $\Pi$. Some simplification for our calculations comes from that such density dependence needs to be accounted in the long wave length limit, $\Pi(q\rightarrow 0,0)$: while the static Lindhard function,  $\Pi_0(q<2k_F,0)=m_{l0}/\pi\hbar^2$ is rigorously constant for all wave vectors up to $2k_F$, the divergence of $V(q)$ at $q\rightarrow 0$, indicates that screening is dominated by the small-$q$ contributions. To describe the density dependence of $\Pi(q\rightarrow 0,0)$, Hwang and Das Sarma~\cite{DasSarma2015} proposed to expand it into a power series in $r_s = \frac{m_{l0} e^2}{\varepsilon \hbar^2 \sqrt{\pi n}}$, the Wigner-Seitz radius (here, $m_{l0}$ is used to avoid double-counting of perturbation theory diagrams), and they have explored the first term in such an expansion. Here, we replicate their approach~\cite{DasSarma2015}, but extend such a calculation onto the second order in $r_s$:
	\begin{equation}
		\Pi (q\rightarrow 0,0) \approx \frac{m_{l0}}{\pi\hbar^2} \left(1 + \aleph_1 r_s + \aleph_2 r_s^2\right),
		\label{eqpseries}
	\end{equation}
	For this, we sum a standard set~\cite{PhysRevB.68.155113,PhysRevB.88.195405,PhysRevB.81.245102,PhysRevB.109.115156} of diagrams drawn in Fig.~\ref{figalldiags}a. A table inset in Fig.~\ref{figalldiags}b summarises the computed values of those 1$^{\rm{st}}$-order ($\Pi_1$) and 2$^{\rm{nd}}$-order ($\Pi_2=\Pi_2^{(1)}+\Pi_2^{(2)}$) contributions, where
	$\Pi_2^{(1)}$ and $\Pi_2^{(2)}$ correspond to one- and two-loop diagrams in Fig.~\ref{figalldiags}a, respectively. The numerical integration of those diagrams produces $\aleph_1=0.29$, $\aleph_2=-0.08$.    
	The opposite signs of the obtained 1st and 2nd order contributions hints at an alternating-sign series. Therefore, we consider the 1$^{\rm{st}}$-order, $\Pi_{\rm{max}}$, and the 2$^{\rm{nd}}$-order, $\Pi_{\rm{min}}$, results as the respective upper and lower bounds for the actual polarisation bubble values, and use their median, $\Pi_*=(\Pi_{\rm{max}}+\Pi_{\rm{min}})/{2}$, to relate the threshold density $n_*$ to $\Delta_0$:
	\begin{align}
		\label{eqpimedian}
		&\Pi_{{\rm max}}=\frac{m_{l0}}{\pi\hbar^2}\left(1+0.29r_s\right);
		\nonumber\\
		&\Pi_{\rm{min}}=\frac{m_{l0}}{\pi\hbar^2}\left(1+0.29r_s-0.08r_s^2\right); 
		\nonumber\\
		&\Pi_* = \frac{m_{l0}}{\pi\hbar^2}\left(1+0.29r_s-0.04r_s^2\right).
	\end{align} 
	This procedure produces the dependence shown in Fig.~\ref{fign1n2}d, where the effect of the SOS enhancement is demonstrated by comparing those values with threshold densities computed with bare SOS. 
	
	At this point, we also need to comment on that the experimentally measured mass (also used in Eq. (\ref{eqnstar})) is quite heavier than the typically DFT-computed value for MoS$_2$~\cite{Kormányos_2015,Kormanyos2013_MoS2_TW,PhysRevB.88.245436,Cheiwchanchamnangij2013_StrainSOC,RodriguesPela2024}. The earlier studies of mass renormalization in 2D semiconductors~\cite{RevModPhys.54.437} and some recent  studies \cite{PhysRevB.69.125334,PhysRevB.70.035104,3wjg-y6qk,PhysRevB.70.035111} suggested a substantial mass enhancement in 2D systems with $r_s \gg 1$. To estimate mass enhancement for electrons in MoS$_2$ at densities around $n_*$, where $r_s \sim 5$, we use the on-shell approximation suggested in Refs.~\cite{PhysRevB.69.125334,PhysRevB.70.035104,PhysRevB.70.035111},
	\begin{equation*}
		\frac{m_{l0}}{m_l}
		= 1 + \frac{m_{l0}}{\hbar^2 k_F}
		\left.
		\frac{d}{dk}\,
		\mathrm{Re}\,\Sigma\!\left(
		k,\,
		\frac{\hbar^2\left(k^2 - k_F^2\right)}{2m_{l0}}
		\right)
		\right|_{k = k_F}.
		\label{eqmassrenorm}
	\end{equation*}
	For this, dynamical polarisability $\Pi(q,\omega)$ is required, which is rather difficult to evaluate within the expansion in $r_s$. In practice, we retain both momentum- and frequency-dependence of $\Pi_0(q,\omega)$ (see Supplemental Information~\cite{supp}), but approximate the higher-order corrections by constant terms already used in the SOS renormalisation analysis. Figure~\ref{figalldiags}b displays mass renormalisation estimated using $\Pi(q,\omega) = \Pi_0(q,\omega) + 0.29 r_s\frac{\pi\hbar^2}{m_{l0}}$, (blue dashed line), $\Pi(q,\omega) = \Pi_0(q,\omega) +(0.29 r_s -0.08  r_s^2)\frac{\pi\hbar^2}{m_{l0}}$, (red dashed line), and the median, $\Pi(q,\omega) = \Pi_0(q,\omega) +(0.29 r_s -0.04  r_s^2)\frac{\pi\hbar^2}{m_{l0}}$, (solid red line).
	For $r_s=5$, all these curves show mass enhancement 35\%-45\%, close enough to the experimentally inferred 50\% enhancement. 

	Finally, we use the calculated SOS renormalisation with the measured $n_*$ and $m_l$ values in Eq.~(\ref{eqnstar}) and determine bare SOS value, $\Delta_0\approx 8.1$ meV, which remains substantially larger than all the existing DFT predictions~\cite{Kormányos_2015,PhysRevB.88.245436,PhysRevB.93.121107,Gjerding_2021,Campi2023}, despite the already incorporated exchange enhancement that reaches the maximum value at the threshold density\footnote{This value can be compared with optical data on hBN-encapsulated MoS$_2$ monolayers~\cite{Robert2020}. Unlike magnetotransport, where many-body interactions renormalize effective masses and SOS, optical spectroscopy probes excitons without free carriers and thus reflects nearly unrenormalized single-particle parameters. Using the framework of Ref.~\cite{PhysRevB.95.081301} and the code of Ref.~\cite{PhysRevB.96.075431}, we recalculate the bare $\Delta_0$ needed to reproduce the reported bright–dark exciton splitting $\Delta_{BD}=14$~meV~\cite{Robert2020}, which depends on  
		\begin{equation}
			\Delta_{BD} = -\Delta_0 + \Delta_{\rm bind} + \Delta_{\rm exch},
		\end{equation}  
		with $\Delta_{\rm bind}$ determined by the effective masses $m_l$, $m_u$, and $m_h$, and $\Delta_{\rm exch}$ the short-range exciton exchange. Taking $m_h \approx 0.6 m_e$~\cite{Eknapakul2014}, we obtain $\Delta_{\rm bind} \approx 19.5$~meV and $\Delta_{\rm exch} \approx 2.6$~meV.}. This suggests that fine tuning of the DFT approach is needed in order to resolve the remaining discrepancy. 
	
	\section{DFT+U+V recipe refined to reproduce the extracted bare SOS values}
	\label{secDFT}
\begin{figure}
		\centering
		\includegraphics[width=0.48\textwidth]{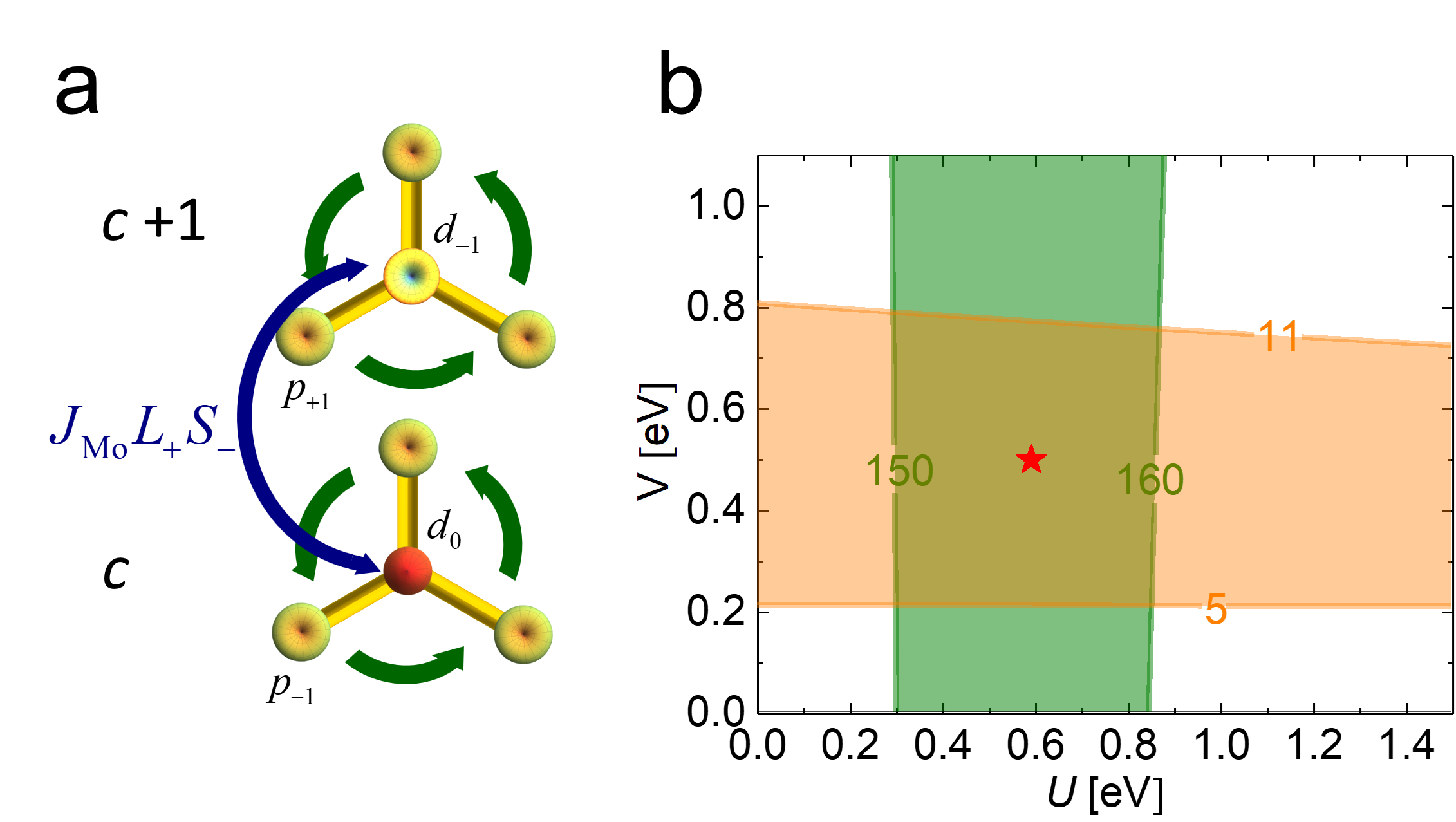}
		\caption{(a) Sketch illustrating how SOS is generated for electrons at the K-valley conduction band edge of MoS$_2$ monolayer.  
			The contributing molybdenum $d$ and sulphur $p$ orbitals are indicated together with their out-of-plane angular momentum projections. Green arrows represent the kinematic angular momentum $\kappa_z$ arising from the Bloch phase factor.  
			Intralayer coupling between conduction and upper conduction band  (denoted as c and c+1, respectively) arises from the part of the atomic spin–orbit operator, $J_{\mathrm{Mo}} L_- S_+$, where $J_{\mathrm{Mo}}$ is the atomic spin–orbit constant of the molybdenum atom.
			(b) Hubbard $U$ and $V$ parameters yielding the range of the conduction-band SOS, $\Delta_0$, (orange area) and valence band SOS, $\Delta_{v}$, (green area).  
			Red star marks the optimal parameters $U_*=0.59$ eV and $V_*=0.5$ eV, corresponding to SOS $\Delta_{0} = 8.1$ meV, $\Delta_{v} = 155$ meV. 
		}
		\label{figUV}
	\end{figure}

	To identify how to tune DFT modelling to change its predictions for SOS at the K-point conduction band edge, we examine the underlying microscopic SOS mechanisms, as illustrated by a sketch in Fig.~\ref{figUV}a. As those MoS$_2$ conduction band edges are dominated by molybdenum $d_{z^2}$ orbitals~\cite{Kormányos_2015,C4CS00301B}, which have zero out-of-plane angular momentum projection ($L_z=0$), the atomic SOC of the transition metal itself does not directly produce the SOS in the band. Instead, it is composed of two contributions (see Supplemental Information~\cite{supp} for details): ($i$) a small admixture of sulphur $p_x$ and $p_y$ orbitals, which carry orbital angular momentum projection $L_z=\mp 1$, brings sulphur's SOS into the conduction band dispersion, set by a spin-orbit constant, $J_{\mathrm{S}}$; and ($ii$) second-order perturbative mixing by the off-diagonal component of atomic SOC in molybdenum, $J_{\mathrm{Mo}}$, between $c$-band's $d_{z^2}$ orbitals and $d_{xz}$ and $d_{yz}$ orbitals ($L_z=\mp 1$) that dominate the $c+1$ conduction band at $\pm K$, as illustrated in Fig.~\ref{figUV}a. Together, these two mechanisms give~\cite{supp},
	\begin{equation}
		\Delta_{\mathrm{0}} = J_{\mathrm{S}} w_{\mathrm{S}}^{c}
		- \frac{3}{2}  \frac{J_{\mathrm{Mo}}^{2} w_{\mathrm{Mo}}^{c} w_{\mathrm{Mo}}^{c+1}}{E_{c+1} - E_c},
		\label{eqsoc}
	\end{equation}
	where $w_{\mathrm{S}}^{c}$ ($w_{\mathrm{Mo}}^{c+1}$) and $E_c$ ($E_{c+1}$) are the weights and band-edge energies of the relevant $p$-orbitals of sulphur in the $c$ band (Mo $d_{xz},d_{yz}$-orbitals in the $c+1$ band), respectively. As atomic SOC constant $J_{\rm{S}}>0$, the two terms in Eq.~(\ref{eqsoc}) set opposite trends, suggesting that they accidentally cancel each other in the earlier implemented DFT computations producing $\Delta_0 \approx 3$ meV. As the balance of such a cancellation is sensitive to the orbital composition of $c$ and $c+1$ bands, we argue that a very small, SOS $\sim 3$ meV, emerged in those DFT studies of MoS$_2$, arises from an underestimated Mo-S hybridisation. Furthermore, the above discussion can be extended~\cite{supp} onto the SOS in the top monolayer in an electrically biased few-layer films studied in Refs~\cite{PhysRevLett.123.117702,PhysRevResearch.5.013113,PhysRevResearch.3.023047,Masseroni2024,Masseroni2025}, which shows that the interlayer hybridisation of orbitals, described in Supplemental Information~\cite{supp}, can produce a small reduction in the value of the on-layer SOS. 
	
	To tune the orbital structure of the bands and promote Mo-S hybridisation towards obtaining bare SOS in the range of $\Delta_0\sim 10$ meV, we employ the DFT+$U$+$V$ approach~\cite{campo2010extended}, which includes the inter-site (closest neighbour) Hubbard interaction $V$ of electrons in orbitals belonging to metal and chalcogen sites, and perform DFT+$U$+$V$ calculations using the Quantum ESPRESSO code. A $31 \times 31 \times 1$ Monkhorst-Pack \textit{k}-point grid is used to sample the Brillouin zone, with plane-wave and charge density cutoffs of 80 and 800 eV, respectively, and a Fermi-Dirac smearing of the electronic distribution of width $\sigma = 0.01$ eV. Optimised fully-relativistic Vanderbilt norm-conserving pseudopotentials were used to approximate electron-ion interactions and to include spin-orbit coupling in all calculations. 
	
	Figure \ref{figUV}b shows the computed $U,V$-dependence of the conduction-band SOS, $\Delta_{0}$, together with the valence-band SOS, $\Delta_{v}$. The computed data are mapped onto the plane of input parameters, $U$ and $V$. Here, the green-painted area 
	identifies the range where the values of the DFT+U+V computed valence-band SOS, $\Delta_v$, fall within the range set by the earlier published experimental studies of the MoS$_2$ valence band properties~\cite{Mak2012,Klots2014,PhysRevB.90.195434,Kozawa2014,doi:10.1021/acs.jpclett.3c02431};
	the orange-painted area covers the range that yields $5\rm{\, meV}<\Delta_0 <11\rm{\, meV}$, chosen to fit between the values of bare SOS extracted from the experimentally measured $n_*$ using $\Pi_{\rm{min}}<\Pi<\Pi_{\rm{max}}$ in Eq.~\ref{eqpimedian}. The red star marks the optimal choice, $U_*=0.59$ eV and $V_*=0.5$ eV, that gives $\Delta_{0}=8.1$ meV, as obtained from the measured $n_*$ using the median $\Pi_*$, and $\Delta_{v}=155$ meV. To mention, the computed effective conduction band masses are almost independent of $U$ and $V$ across the analysed range (e.g., $m_{u0}=0.485m_e$ and $m_{l0}=0.433m_e$ for $U_*=0.59$ eV and $V_*=0.5$ eV).
	
	We also note that the implementation of DFT+U+V with $U_*=0.59$ eV and $V_*=0.5$ eV significantly improves the agreement between the computed band gap, which increases to $E_g = 1.90$ eV (as compared to $\approx1.7$ eV at $U=V=0$), and the bounds~\cite{PhysRevLett.51.1884} set by the experimentally established optical ($\approx 1.80$ eV) and quasiparticle ($\approx 2.16$ eV) monolayer band gaps~\cite{PhysRevB.111.075423,PhysRevLett.105.136805}. Overall, the obtained parameters support a picture of unconventional screening, combining weak short-range and stronger long-range Hubbard interactions~\cite{PhysRevB.109.125108}.

	\section{Summary}
	Based on the analysis of Shubnikov de Haas oscillations, observed in monolayer MoS$_2$, we found the threshold density, $n_*$, for the onset of filling the states in the upper spin-orbit split (SOS) conduction band in this material, which appeared to be much higher than the value anticipated based on the earlier DFT-computed band parameters of MoS$_2$. While the doping-induced (hence, doping-dependent) exchange enhancement of SOS and mass enhancement partly explain the discrepancy between DFT results and the measured density threshold, theory and experiment for the conduction-band spin–orbit splitting in monolayer MoS$_2$, the estimated value of bare SOS, $\Delta_0 \approx 8.1$~meV, required to obtain the measured threshold density is still much higher than the previously DFT-computed values. However, the discrepancy can be resolved by using a compact DFT+$U$+$V$ computational scheme that simultaneously corrects the conduction- and valence-band SOS, while improving the prediction for the quasiparticle band gap: we argue that conventional DFT approach severely underestimates the conduction-band SOS due to an accidental cancellation of orbital contributions when correlations are neglected. Including on-site $U_*=0.59$ eV and inter-site $V_*=0.5$ eV Hubbard interaction provides the value of $\Delta_0$ extracted from the experiment, as well as produces the valence-band SOS and the monolayer band gap values closer to the experimentally established ones. The resulting approach offers a computationally efficient route to accurate spin--orbit and band-edge parameters across the TMDs family without resorting to resource-hungry GW methods.

	\section{Acknowledgements}
	We thank A. Principi, Y. Gindikin, L. Golub and M. Potemski for valuable discussions. We acknowledge support from 2DFERROPLEX project funded by the European Union under grant agreement 101223116, 
	EPSRC Grants EP/S019367/1, EP/P026850/1, and EP/N010345/1; British Council project 1185409051.
	X.L. acknowledges financial support from the University of Manchester’s Dean’s Doctoral Scholarship. 
	S.A. acknowledges the support of the Deanship
	of Scientfic Research at Northern Border University, Arar, KSA for funding through the project number NBU-SA-FIR-2025,  
	J.G.M. is supported by the University of Manchester Dame Kathleen Ollerenshaw Fellowship. The authors acknowledge the use of resources provided by the Isambard 3 Tier-2 HPC Facility, hosted by the University of Bristol and operated by the GW4 Alliance (\url{https://gw4.ac.uk}) and funded by UK Research and Innovation; and the Engineering and Physical Sciences Research Council [EP/X039137/1]; and the Sulis Tier 2 HPC platform hosted by the Scientific Computing Research Technology Platform at the University of Warwick, funded by EPSRC Grant EP/T022108/1 and the HPC Midlands+ consortium.
	
	\section{Contributions}
	This was conceived by VF, IR, and KE. Experimental data were obtained and analysed by MM, RP, TI, and KE. IR and VF modelled exchange SOS enhancement. DFT modelling and its tuning was performed by XL and SA. The DFT+U+V model was set by JM and VF and realised by XL and JM. The paper has been written by IR, JM, MM and VF with contributions from all authors.
	
	\bibliography{references}

\begin{thebibliography}{79}%
\makeatletter
\providecommand \@ifxundefined [1]{%
 \@ifx{#1\undefined}
}%
\providecommand \@ifnum [1]{%
 \ifnum #1\expandafter \@firstoftwo
 \else \expandafter \@secondoftwo
 \fi
}%
\providecommand \@ifx [1]{%
 \ifx #1\expandafter \@firstoftwo
 \else \expandafter \@secondoftwo
 \fi
}%
\providecommand \natexlab [1]{#1}%
\providecommand \enquote  [1]{``#1''}%
\providecommand \bibnamefont  [1]{#1}%
\providecommand \bibfnamefont [1]{#1}%
\providecommand \citenamefont [1]{#1}%
\providecommand \href@noop [0]{\@secondoftwo}%
\providecommand \href [0]{\begingroup \@sanitize@url \@href}%
\providecommand \@href[1]{\@@startlink{#1}\@@href}%
\providecommand \@@href[1]{\endgroup#1\@@endlink}%
\providecommand \@sanitize@url [0]{\catcode `\\12\catcode `\$12\catcode
  `\&12\catcode `\#12\catcode `\^12\catcode `\_12\catcode `\%12\relax}%
\providecommand \@@startlink[1]{}%
\providecommand \@@endlink[0]{}%
\providecommand \url  [0]{\begingroup\@sanitize@url \@url }%
\providecommand \@url [1]{\endgroup\@href {#1}{\urlprefix }}%
\providecommand \urlprefix  [0]{URL }%
\providecommand \Eprint [0]{\href }%
\providecommand \doibase [0]{https://doi.org/}%
\providecommand \selectlanguage [0]{\@gobble}%
\providecommand \bibinfo  [0]{\@secondoftwo}%
\providecommand \bibfield  [0]{\@secondoftwo}%
\providecommand \translation [1]{[#1]}%
\providecommand \BibitemOpen [0]{}%
\providecommand \bibitemStop [0]{}%
\providecommand \bibitemNoStop [0]{.\EOS\space}%
\providecommand \EOS [0]{\spacefactor3000\relax}%
\providecommand \BibitemShut  [1]{\csname bibitem#1\endcsname}%
\let\auto@bib@innerbib\@empty
%</preamble>
\bibitem [{\citenamefont {Radisavljevic}\ and\ \citenamefont
  {Kis}(2013)}]{Radisavljevic2013MobilityEngineeringMoS2}%
  \BibitemOpen
  \bibfield  {author} {\bibinfo {author} {\bibfnamefont {B.}~\bibnamefont
  {Radisavljevic}}\ and\ \bibinfo {author} {\bibfnamefont {A.}~\bibnamefont
  {Kis}},\ }\bibfield  {title} {\bibinfo {title} {Mobility engineering and a
  metal-insulator transition in monolayer mos2},\ }\href
  {https://doi.org/10.1038/nmat3687} {\bibfield  {journal} {\bibinfo  {journal}
  {Nature Materials}\ }\textbf {\bibinfo {volume} {12}},\ \bibinfo {pages}
  {815} (\bibinfo {year} {2013})}\BibitemShut {NoStop}%
\bibitem [{\citenamefont {Lembke}\ \emph {et~al.}(2015)\citenamefont {Lembke},
  \citenamefont {Bertolazzi},\ and\ \citenamefont
  {Kis}}]{Lembke2015SingleLayerMoS2Electronics}%
  \BibitemOpen
  \bibfield  {author} {\bibinfo {author} {\bibfnamefont {D.}~\bibnamefont
  {Lembke}}, \bibinfo {author} {\bibfnamefont {S.}~\bibnamefont {Bertolazzi}},\
  and\ \bibinfo {author} {\bibfnamefont {A.}~\bibnamefont {Kis}},\ }\bibfield
  {title} {\bibinfo {title} {Single-layer mos2 electronics},\ }\href
  {https://doi.org/10.1021/ar500274q} {\bibfield  {journal} {\bibinfo
  {journal} {Accounts of Chemical Research}\ }\textbf {\bibinfo {volume}
  {48}},\ \bibinfo {pages} {100} (\bibinfo {year} {2015})}\BibitemShut
  {NoStop}%
\bibitem [{\citenamefont {Dumcenco}\ \emph {et~al.}(2015)\citenamefont
  {Dumcenco}, \citenamefont {Ovchinnikov}, \citenamefont {Marinov},
  \citenamefont {Lopez-Sanchez}, \citenamefont {Krasnozhon}, \citenamefont
  {Chen}, \citenamefont {Gillet}, \citenamefont {i~Morral}, \citenamefont
  {Radenovic},\ and\ \citenamefont {Kis}}]{Dumcenco2015LargeAreaMoS2}%
  \BibitemOpen
  \bibfield  {author} {\bibinfo {author} {\bibfnamefont {D.}~\bibnamefont
  {Dumcenco}}, \bibinfo {author} {\bibfnamefont {D.}~\bibnamefont
  {Ovchinnikov}}, \bibinfo {author} {\bibfnamefont {K.}~\bibnamefont
  {Marinov}}, \bibinfo {author} {\bibfnamefont {O.}~\bibnamefont
  {Lopez-Sanchez}}, \bibinfo {author} {\bibfnamefont {D.}~\bibnamefont
  {Krasnozhon}}, \bibinfo {author} {\bibfnamefont {M.-W.}\ \bibnamefont
  {Chen}}, \bibinfo {author} {\bibfnamefont {P.}~\bibnamefont {Gillet}},
  \bibinfo {author} {\bibfnamefont {A.~F.}\ \bibnamefont {i~Morral}}, \bibinfo
  {author} {\bibfnamefont {A.}~\bibnamefont {Radenovic}},\ and\ \bibinfo
  {author} {\bibfnamefont {A.}~\bibnamefont {Kis}},\ }\bibfield  {title}
  {\bibinfo {title} {Large-area epitaxial monolayer mos2},\ }\href
  {https://doi.org/10.1021/acsnano.5b01281} {\bibfield  {journal} {\bibinfo
  {journal} {ACS Nano}\ }\textbf {\bibinfo {volume} {9}},\ \bibinfo {pages}
  {4611} (\bibinfo {year} {2015})}\BibitemShut {NoStop}%
\bibitem [{\citenamefont {Withers}\ \emph
  {et~al.}(2015{\natexlab{a}})\citenamefont {Withers}, \citenamefont {Del
  Pozo-Zamudio}, \citenamefont {Mishchenko}, \citenamefont {Rooney},
  \citenamefont {Gholinia}, \citenamefont {Watanabe}, \citenamefont
  {Taniguchi}, \citenamefont {Haigh}, \citenamefont {Geim}, \citenamefont
  {Tartakovskii},\ and\ \citenamefont {Novoselov}}]{Withers2015}%
  \BibitemOpen
  \bibfield  {author} {\bibinfo {author} {\bibfnamefont {F.}~\bibnamefont
  {Withers}}, \bibinfo {author} {\bibfnamefont {O.}~\bibnamefont {Del
  Pozo-Zamudio}}, \bibinfo {author} {\bibfnamefont {A.}~\bibnamefont
  {Mishchenko}}, \bibinfo {author} {\bibfnamefont {A.~P.}\ \bibnamefont
  {Rooney}}, \bibinfo {author} {\bibfnamefont {A.}~\bibnamefont {Gholinia}},
  \bibinfo {author} {\bibfnamefont {K.}~\bibnamefont {Watanabe}}, \bibinfo
  {author} {\bibfnamefont {T.}~\bibnamefont {Taniguchi}}, \bibinfo {author}
  {\bibfnamefont {S.~J.}\ \bibnamefont {Haigh}}, \bibinfo {author}
  {\bibfnamefont {A.~K.}\ \bibnamefont {Geim}}, \bibinfo {author}
  {\bibfnamefont {A.~I.}\ \bibnamefont {Tartakovskii}},\ and\ \bibinfo {author}
  {\bibfnamefont {K.~S.}\ \bibnamefont {Novoselov}},\ }\bibfield  {title}
  {\bibinfo {title} {Light-emitting diodes by band-structure engineering in van
  der waals heterostructures},\ }\href {https://doi.org/10.1038/nmat4205}
  {\bibfield  {journal} {\bibinfo  {journal} {Nature Materials}\ }\textbf
  {\bibinfo {volume} {14}},\ \bibinfo {pages} {301} (\bibinfo {year}
  {2015}{\natexlab{a}})}\BibitemShut {NoStop}%
\bibitem [{\citenamefont {Withers}\ \emph
  {et~al.}(2015{\natexlab{b}})\citenamefont {Withers}, \citenamefont {Del
  Pozo-Zamudio}, \citenamefont {Schwarz}, \citenamefont {Dufferwiel},
  \citenamefont {Walker}, \citenamefont {Godde}, \citenamefont {Rooney},
  \citenamefont {Gholinia}, \citenamefont {Woods}, \citenamefont {Blake},
  \citenamefont {Haigh}, \citenamefont {Watanabe}, \citenamefont {Taniguchi},
  \citenamefont {Aleiner}, \citenamefont {Geim}, \citenamefont {Fal'ko},
  \citenamefont {Tartakovskii},\ and\ \citenamefont
  {Novoselov}}]{Withers2015_2}%
  \BibitemOpen
  \bibfield  {author} {\bibinfo {author} {\bibfnamefont {F.}~\bibnamefont
  {Withers}}, \bibinfo {author} {\bibfnamefont {O.}~\bibnamefont {Del
  Pozo-Zamudio}}, \bibinfo {author} {\bibfnamefont {S.}~\bibnamefont
  {Schwarz}}, \bibinfo {author} {\bibfnamefont {S.}~\bibnamefont {Dufferwiel}},
  \bibinfo {author} {\bibfnamefont {P.~M.}\ \bibnamefont {Walker}}, \bibinfo
  {author} {\bibfnamefont {T.}~\bibnamefont {Godde}}, \bibinfo {author}
  {\bibfnamefont {A.~P.}\ \bibnamefont {Rooney}}, \bibinfo {author}
  {\bibfnamefont {A.}~\bibnamefont {Gholinia}}, \bibinfo {author}
  {\bibfnamefont {C.~R.}\ \bibnamefont {Woods}}, \bibinfo {author}
  {\bibfnamefont {P.}~\bibnamefont {Blake}}, \bibinfo {author} {\bibfnamefont
  {S.~J.}\ \bibnamefont {Haigh}}, \bibinfo {author} {\bibfnamefont
  {K.}~\bibnamefont {Watanabe}}, \bibinfo {author} {\bibfnamefont
  {T.}~\bibnamefont {Taniguchi}}, \bibinfo {author} {\bibfnamefont {I.~L.}\
  \bibnamefont {Aleiner}}, \bibinfo {author} {\bibfnamefont {A.~K.}\
  \bibnamefont {Geim}}, \bibinfo {author} {\bibfnamefont {V.~I.}\ \bibnamefont
  {Fal'ko}}, \bibinfo {author} {\bibfnamefont {A.~I.}\ \bibnamefont
  {Tartakovskii}},\ and\ \bibinfo {author} {\bibfnamefont {K.~S.}\ \bibnamefont
  {Novoselov}},\ }\bibfield  {title} {\bibinfo {title} {Wse2 light-emitting
  tunneling transistors with enhanced brightness at room temperature},\ }\href
  {https://doi.org/10.1021/acs.nanolett.5b03740} {\bibfield  {journal}
  {\bibinfo  {journal} {Nano Letters}\ }\textbf {\bibinfo {volume} {15}},\
  \bibinfo {pages} {8223} (\bibinfo {year} {2015}{\natexlab{b}})}\BibitemShut
  {NoStop}%
\bibitem [{\citenamefont {Tang}\ \emph {et~al.}(2019)\citenamefont {Tang},
  \citenamefont {Mak},\ and\ \citenamefont {Shan}}]{Tang2019}%
  \BibitemOpen
  \bibfield  {author} {\bibinfo {author} {\bibfnamefont {Y.}~\bibnamefont
  {Tang}}, \bibinfo {author} {\bibfnamefont {K.~F.}\ \bibnamefont {Mak}},\ and\
  \bibinfo {author} {\bibfnamefont {J.}~\bibnamefont {Shan}},\ }\bibfield
  {title} {\bibinfo {title} {{Long valley lifetime of dark excitons in
  single-layer WSe2}},\ }\href {https://doi.org/10.1038/s41467-019-12129-1}
  {\bibfield  {journal} {\bibinfo  {journal} {Nature Communications}\ }\textbf
  {\bibinfo {volume} {10}},\ \bibinfo {pages} {4047} (\bibinfo {year}
  {2019})}\BibitemShut {NoStop}%
\bibitem [{\citenamefont {Dey}\ \emph {et~al.}(2017)\citenamefont {Dey},
  \citenamefont {Yang}, \citenamefont {Robert}, \citenamefont {Wang},
  \citenamefont {Urbaszek}, \citenamefont {Marie},\ and\ \citenamefont
  {Crooker}}]{PhysRevLett.119.137401}%
  \BibitemOpen
  \bibfield  {author} {\bibinfo {author} {\bibfnamefont {P.}~\bibnamefont
  {Dey}}, \bibinfo {author} {\bibfnamefont {L.}~\bibnamefont {Yang}}, \bibinfo
  {author} {\bibfnamefont {C.}~\bibnamefont {Robert}}, \bibinfo {author}
  {\bibfnamefont {G.}~\bibnamefont {Wang}}, \bibinfo {author} {\bibfnamefont
  {B.}~\bibnamefont {Urbaszek}}, \bibinfo {author} {\bibfnamefont
  {X.}~\bibnamefont {Marie}},\ and\ \bibinfo {author} {\bibfnamefont {S.~A.}\
  \bibnamefont {Crooker}},\ }\bibfield  {title} {\bibinfo {title}
  {{Gate-Controlled Spin-Valley Locking of Resident Carriers in
  ${\mathrm{WSe}}_{2}$ Monolayers}},\ }\href
  {https://doi.org/10.1103/PhysRevLett.119.137401} {\bibfield  {journal}
  {\bibinfo  {journal} {Phys. Rev. Lett.}\ }\textbf {\bibinfo {volume} {119}},\
  \bibinfo {pages} {137401} (\bibinfo {year} {2017})}\BibitemShut {NoStop}%
\bibitem [{\citenamefont {Perkins}\ and\ \citenamefont
  {Ferreira}(2024)}]{PERKINS2024205}%
  \BibitemOpen
  \bibfield  {author} {\bibinfo {author} {\bibfnamefont {D.~T.}\ \bibnamefont
  {Perkins}}\ and\ \bibinfo {author} {\bibfnamefont {A.}~\bibnamefont
  {Ferreira}},\ }\bibfield  {title} {\bibinfo {title} {Spintronics in 2d
  graphene-based van der waals heterostructures},\ }in\ \href
  {https://doi.org/https://doi.org/10.1016/B978-0-323-90800-9.00203-1} {\emph
  {\bibinfo {booktitle} {Encyclopedia of Condensed Matter Physics (Second
  Edition)}}},\ \bibinfo {editor} {edited by\ \bibinfo {editor} {\bibfnamefont
  {T.}~\bibnamefont {Chakraborty}}}\ (\bibinfo  {publisher} {Academic Press},\
  \bibinfo {address} {Oxford},\ \bibinfo {year} {2024})\ \bibinfo {edition}
  {second edition}\ ed.,\ pp.\ \bibinfo {pages} {205--222}\BibitemShut
  {NoStop}%
\bibitem [{\citenamefont {Ahn}(2020)}]{Ahn2020}%
  \BibitemOpen
  \bibfield  {author} {\bibinfo {author} {\bibfnamefont {E.~C.}\ \bibnamefont
  {Ahn}},\ }\bibfield  {title} {\bibinfo {title} {2d materials for spintronic
  devices},\ }\href {https://doi.org/10.1038/s41699-020-0152-0} {\bibfield
  {journal} {\bibinfo  {journal} {npj 2D Materials and Applications}\ }\textbf
  {\bibinfo {volume} {4}},\ \bibinfo {pages} {17} (\bibinfo {year}
  {2020})}\BibitemShut {NoStop}%
\bibitem [{\citenamefont {Manzeli}\ \emph {et~al.}(2017)\citenamefont
  {Manzeli}, \citenamefont {Ovchinnikov}, \citenamefont {Pasquier},
  \citenamefont {Yazyev},\ and\ \citenamefont {Kis}}]{Manzeli2017}%
  \BibitemOpen
  \bibfield  {author} {\bibinfo {author} {\bibfnamefont {S.}~\bibnamefont
  {Manzeli}}, \bibinfo {author} {\bibfnamefont {D.}~\bibnamefont
  {Ovchinnikov}}, \bibinfo {author} {\bibfnamefont {D.}~\bibnamefont
  {Pasquier}}, \bibinfo {author} {\bibfnamefont {O.~V.}\ \bibnamefont
  {Yazyev}},\ and\ \bibinfo {author} {\bibfnamefont {A.}~\bibnamefont {Kis}},\
  }\bibfield  {title} {\bibinfo {title} {2d transition metal dichalcogenides},\
  }\href {https://doi.org/10.1038/natrevmats.2017.33} {\bibfield  {journal}
  {\bibinfo  {journal} {Nature Reviews Materials}\ }\textbf {\bibinfo {volume}
  {2}},\ \bibinfo {pages} {17033} (\bibinfo {year} {2017})}\BibitemShut
  {NoStop}%
\bibitem [{\citenamefont {Regan}\ \emph {et~al.}(2022)\citenamefont {Regan},
  \citenamefont {Wang}, \citenamefont {Paik}, \citenamefont {Zeng},
  \citenamefont {Zhang}, \citenamefont {Zhu}, \citenamefont {MacDonald},
  \citenamefont {Deng},\ and\ \citenamefont {Wang}}]{Regan2022}%
  \BibitemOpen
  \bibfield  {author} {\bibinfo {author} {\bibfnamefont {E.~C.}\ \bibnamefont
  {Regan}}, \bibinfo {author} {\bibfnamefont {D.}~\bibnamefont {Wang}},
  \bibinfo {author} {\bibfnamefont {E.~Y.}\ \bibnamefont {Paik}}, \bibinfo
  {author} {\bibfnamefont {Y.}~\bibnamefont {Zeng}}, \bibinfo {author}
  {\bibfnamefont {L.}~\bibnamefont {Zhang}}, \bibinfo {author} {\bibfnamefont
  {J.}~\bibnamefont {Zhu}}, \bibinfo {author} {\bibfnamefont {A.~H.}\
  \bibnamefont {MacDonald}}, \bibinfo {author} {\bibfnamefont {H.}~\bibnamefont
  {Deng}},\ and\ \bibinfo {author} {\bibfnamefont {F.}~\bibnamefont {Wang}},\
  }\bibfield  {title} {\bibinfo {title} {Emerging exciton physics in transition
  metal dichalcogenide heterobilayers},\ }\href
  {https://doi.org/10.1038/s41578-022-00440-1} {\bibfield  {journal} {\bibinfo
  {journal} {Nature Reviews Materials}\ }\textbf {\bibinfo {volume} {7}},\
  \bibinfo {pages} {778} (\bibinfo {year} {2022})}\BibitemShut {NoStop}%
\bibitem [{\citenamefont {Wang}\ \emph {et~al.}(2018)\citenamefont {Wang},
  \citenamefont {Chernikov}, \citenamefont {Glazov}, \citenamefont {Heinz},
  \citenamefont {Marie}, \citenamefont {Amand},\ and\ \citenamefont
  {Urbaszek}}]{RevModPhys.90.021001}%
  \BibitemOpen
  \bibfield  {author} {\bibinfo {author} {\bibfnamefont {G.}~\bibnamefont
  {Wang}}, \bibinfo {author} {\bibfnamefont {A.}~\bibnamefont {Chernikov}},
  \bibinfo {author} {\bibfnamefont {M.~M.}\ \bibnamefont {Glazov}}, \bibinfo
  {author} {\bibfnamefont {T.~F.}\ \bibnamefont {Heinz}}, \bibinfo {author}
  {\bibfnamefont {X.}~\bibnamefont {Marie}}, \bibinfo {author} {\bibfnamefont
  {T.}~\bibnamefont {Amand}},\ and\ \bibinfo {author} {\bibfnamefont
  {B.}~\bibnamefont {Urbaszek}},\ }\bibfield  {title} {\bibinfo {title}
  {Colloquium: Excitons in atomically thin transition metal dichalcogenides},\
  }\href {https://doi.org/10.1103/RevModPhys.90.021001} {\bibfield  {journal}
  {\bibinfo  {journal} {Rev. Mod. Phys.}\ }\textbf {\bibinfo {volume} {90}},\
  \bibinfo {pages} {021001} (\bibinfo {year} {2018})}\BibitemShut {NoStop}%
\bibitem [{\citenamefont {O'Brien}\ \emph {et~al.}(2023)\citenamefont
  {O'Brien}, \citenamefont {Naylor}, \citenamefont {Dorow}, \citenamefont
  {Maxey}, \citenamefont {Penumatcha}, \citenamefont {Vyatskikh}, \citenamefont
  {Zhong}, \citenamefont {Kitamura}, \citenamefont {Lee}, \citenamefont
  {Rogan}, \citenamefont {Mortelmans}, \citenamefont {Kavrik}, \citenamefont
  {Steinhardt}, \citenamefont {Buragohain}, \citenamefont {Dutta},
  \citenamefont {Tronic}, \citenamefont {Clendenning}, \citenamefont {Fischer},
  \citenamefont {Putna}, \citenamefont {Radosavljevic}, \citenamefont {Metz},\
  and\ \citenamefont {Avci}}]{OBrien2023}%
  \BibitemOpen
  \bibfield  {author} {\bibinfo {author} {\bibfnamefont {K.~P.}\ \bibnamefont
  {O'Brien}}, \bibinfo {author} {\bibfnamefont {C.~H.}\ \bibnamefont {Naylor}},
  \bibinfo {author} {\bibfnamefont {C.}~\bibnamefont {Dorow}}, \bibinfo
  {author} {\bibfnamefont {K.}~\bibnamefont {Maxey}}, \bibinfo {author}
  {\bibfnamefont {A.~V.}\ \bibnamefont {Penumatcha}}, \bibinfo {author}
  {\bibfnamefont {A.}~\bibnamefont {Vyatskikh}}, \bibinfo {author}
  {\bibfnamefont {T.}~\bibnamefont {Zhong}}, \bibinfo {author} {\bibfnamefont
  {A.}~\bibnamefont {Kitamura}}, \bibinfo {author} {\bibfnamefont
  {S.}~\bibnamefont {Lee}}, \bibinfo {author} {\bibfnamefont {C.}~\bibnamefont
  {Rogan}}, \bibinfo {author} {\bibfnamefont {W.}~\bibnamefont {Mortelmans}},
  \bibinfo {author} {\bibfnamefont {M.~S.}\ \bibnamefont {Kavrik}}, \bibinfo
  {author} {\bibfnamefont {R.}~\bibnamefont {Steinhardt}}, \bibinfo {author}
  {\bibfnamefont {P.}~\bibnamefont {Buragohain}}, \bibinfo {author}
  {\bibfnamefont {S.}~\bibnamefont {Dutta}}, \bibinfo {author} {\bibfnamefont
  {T.}~\bibnamefont {Tronic}}, \bibinfo {author} {\bibfnamefont
  {S.}~\bibnamefont {Clendenning}}, \bibinfo {author} {\bibfnamefont
  {P.}~\bibnamefont {Fischer}}, \bibinfo {author} {\bibfnamefont {E.~S.}\
  \bibnamefont {Putna}}, \bibinfo {author} {\bibfnamefont {M.}~\bibnamefont
  {Radosavljevic}}, \bibinfo {author} {\bibfnamefont {M.}~\bibnamefont
  {Metz}},\ and\ \bibinfo {author} {\bibfnamefont {U.}~\bibnamefont {Avci}},\
  }\bibfield  {title} {\bibinfo {title} {Process integration and future outlook
  of 2d transistors},\ }\href {https://doi.org/10.1038/s41467-023-41779-5}
  {\bibfield  {journal} {\bibinfo  {journal} {Nature Communications}\ }\textbf
  {\bibinfo {volume} {14}},\ \bibinfo {pages} {6400} (\bibinfo {year}
  {2023})}\BibitemShut {NoStop}%
\bibitem [{\citenamefont {Schaibley}\ \emph {et~al.}(2016)\citenamefont
  {Schaibley}, \citenamefont {Yu}, \citenamefont {Clark}, \citenamefont
  {Rivera}, \citenamefont {Ross}, \citenamefont {Seyler}, \citenamefont {Yao},\
  and\ \citenamefont {Xu}}]{Schaibley2016}%
  \BibitemOpen
  \bibfield  {author} {\bibinfo {author} {\bibfnamefont {J.~R.}\ \bibnamefont
  {Schaibley}}, \bibinfo {author} {\bibfnamefont {H.}~\bibnamefont {Yu}},
  \bibinfo {author} {\bibfnamefont {G.}~\bibnamefont {Clark}}, \bibinfo
  {author} {\bibfnamefont {P.}~\bibnamefont {Rivera}}, \bibinfo {author}
  {\bibfnamefont {J.~S.}\ \bibnamefont {Ross}}, \bibinfo {author}
  {\bibfnamefont {K.~L.}\ \bibnamefont {Seyler}}, \bibinfo {author}
  {\bibfnamefont {W.}~\bibnamefont {Yao}},\ and\ \bibinfo {author}
  {\bibfnamefont {X.}~\bibnamefont {Xu}},\ }\bibfield  {title} {\bibinfo
  {title} {Valleytronics in 2d materials},\ }\href
  {https://doi.org/10.1038/natrevmats.2016.55} {\bibfield  {journal} {\bibinfo
  {journal} {Nature Reviews Materials}\ }\textbf {\bibinfo {volume} {1}},\
  \bibinfo {pages} {16055} (\bibinfo {year} {2016})}\BibitemShut {NoStop}%
\bibitem [{\citenamefont {Yang}\ \emph {et~al.}(2023)\citenamefont {Yang},
  \citenamefont {Liu}, \citenamefont {Li}, \citenamefont {Hu}, \citenamefont
  {Zhou}, \citenamefont {Zhu}, \citenamefont {Chen}, \citenamefont {Gao},\ and\
  \citenamefont {Zhang}}]{yang_wafer_scale_mos2_2023}%
  \BibitemOpen
  \bibfield  {author} {\bibinfo {author} {\bibfnamefont {P.}~\bibnamefont
  {Yang}}, \bibinfo {author} {\bibfnamefont {F.}~\bibnamefont {Liu}}, \bibinfo
  {author} {\bibfnamefont {X.}~\bibnamefont {Li}}, \bibinfo {author}
  {\bibfnamefont {J.}~\bibnamefont {Hu}}, \bibinfo {author} {\bibfnamefont
  {F.}~\bibnamefont {Zhou}}, \bibinfo {author} {\bibfnamefont {L.}~\bibnamefont
  {Zhu}}, \bibinfo {author} {\bibfnamefont {Q.}~\bibnamefont {Chen}}, \bibinfo
  {author} {\bibfnamefont {P.}~\bibnamefont {Gao}},\ and\ \bibinfo {author}
  {\bibfnamefont {Y.}~\bibnamefont {Zhang}},\ }\bibfield  {title} {\bibinfo
  {title} {Highly reproducible epitaxial growth of wafer-scale single-crystal
  monolayer {MoS2} on sapphire},\ }\href
  {https://doi.org/10.1002/smtd.202300165} {\bibfield  {journal} {\bibinfo
  {journal} {Small Methods}\ }\textbf {\bibinfo {volume} {7}},\ \bibinfo
  {pages} {e2300165} (\bibinfo {year} {2023})}\BibitemShut {NoStop}%
\bibitem [{\citenamefont {Xia}\ \emph {et~al.}(2023)\citenamefont {Xia},
  \citenamefont {Chen}, \citenamefont {Wei}, \citenamefont {Wang},
  \citenamefont {Chen}, \citenamefont {Zhou} \emph
  {et~al.}}]{xia_12inch_mos2_2023}%
  \BibitemOpen
  \bibfield  {author} {\bibinfo {author} {\bibfnamefont {Y.}~\bibnamefont
  {Xia}}, \bibinfo {author} {\bibfnamefont {X.}~\bibnamefont {Chen}}, \bibinfo
  {author} {\bibfnamefont {J.}~\bibnamefont {Wei}}, \bibinfo {author}
  {\bibfnamefont {S.}~\bibnamefont {Wang}}, \bibinfo {author} {\bibfnamefont
  {S.}~\bibnamefont {Chen}}, \bibinfo {author} {\bibfnamefont {P.}~\bibnamefont
  {Zhou}}, \emph {et~al.},\ }\bibfield  {title} {\bibinfo {title} {12-inch
  growth of uniform {MoS2} monolayer for integrated circuit manufacture},\
  }\href {https://doi.org/10.1038/s41563-023-01671-5} {\bibfield  {journal}
  {\bibinfo  {journal} {Nature Materials}\ }\textbf {\bibinfo {volume} {22}},\
  \bibinfo {pages} {1324} (\bibinfo {year} {2023})}\BibitemShut {NoStop}%
\bibitem [{\citenamefont {Li}\ \emph {et~al.}(2024)\citenamefont {Li},
  \citenamefont {Wang}, \citenamefont {Wu}, \citenamefont {Xu}, \citenamefont
  {Tian}, \citenamefont {Huang}, \citenamefont {Wang}, \citenamefont {Zhao},
  \citenamefont {Zhang}, \citenamefont {Fan} \emph
  {et~al.}}]{li_epitaxy_buffer_layer_2024}%
  \BibitemOpen
  \bibfield  {author} {\bibinfo {author} {\bibfnamefont {L.}~\bibnamefont
  {Li}}, \bibinfo {author} {\bibfnamefont {Q.}~\bibnamefont {Wang}}, \bibinfo
  {author} {\bibfnamefont {F.}~\bibnamefont {Wu}}, \bibinfo {author}
  {\bibfnamefont {Q.}~\bibnamefont {Xu}}, \bibinfo {author} {\bibfnamefont
  {J.}~\bibnamefont {Tian}}, \bibinfo {author} {\bibfnamefont {Z.}~\bibnamefont
  {Huang}}, \bibinfo {author} {\bibfnamefont {Q.}~\bibnamefont {Wang}},
  \bibinfo {author} {\bibfnamefont {X.}~\bibnamefont {Zhao}}, \bibinfo {author}
  {\bibfnamefont {Q.}~\bibnamefont {Zhang}}, \bibinfo {author} {\bibfnamefont
  {Q.}~\bibnamefont {Fan}}, \emph {et~al.},\ }\bibfield  {title} {\bibinfo
  {title} {Epitaxy of wafer-scale single-crystal {MoS2} monolayer via buffer
  layer control},\ }\href {https://doi.org/10.1038/s41467-024-46170-6}
  {\bibfield  {journal} {\bibinfo  {journal} {Nature Communications}\ }\textbf
  {\bibinfo {volume} {15}},\ \bibinfo {pages} {1825} (\bibinfo {year}
  {2024})}\BibitemShut {NoStop}%
\bibitem [{\citenamefont {Yu}\ \emph {et~al.}(2024)\citenamefont {Yu},
  \citenamefont {Huang}, \citenamefont {Zhou}, \citenamefont {Peng},
  \citenamefont {Li}, \citenamefont {Yin}, \citenamefont {Zhao}, \citenamefont
  {Zhu}, \citenamefont {Wang}, \citenamefont {Liu} \emph
  {et~al.}}]{yu_eight_inch_mos2_2024}%
  \BibitemOpen
  \bibfield  {author} {\bibinfo {author} {\bibfnamefont {H.}~\bibnamefont
  {Yu}}, \bibinfo {author} {\bibfnamefont {L.}~\bibnamefont {Huang}}, \bibinfo
  {author} {\bibfnamefont {L.}~\bibnamefont {Zhou}}, \bibinfo {author}
  {\bibfnamefont {Y.}~\bibnamefont {Peng}}, \bibinfo {author} {\bibfnamefont
  {X.}~\bibnamefont {Li}}, \bibinfo {author} {\bibfnamefont {P.}~\bibnamefont
  {Yin}}, \bibinfo {author} {\bibfnamefont {J.}~\bibnamefont {Zhao}}, \bibinfo
  {author} {\bibfnamefont {M.}~\bibnamefont {Zhu}}, \bibinfo {author}
  {\bibfnamefont {S.}~\bibnamefont {Wang}}, \bibinfo {author} {\bibfnamefont
  {J.}~\bibnamefont {Liu}}, \emph {et~al.},\ }\bibfield  {title} {\bibinfo
  {title} {Eight in. wafer-scale epitaxial monolayer {MoS2}},\ }\href
  {https://doi.org/10.1002/adma.202402855} {\bibfield  {journal} {\bibinfo
  {journal} {Advanced Materials}\ }\textbf {\bibinfo {volume} {36}},\ \bibinfo
  {pages} {e2402855} (\bibinfo {year} {2024})}\BibitemShut {NoStop}%
\bibitem [{\citenamefont {Kwon}\ \emph {et~al.}(2024)\citenamefont {Kwon},
  \citenamefont {Seol}, \citenamefont {Yoo}, \citenamefont {Ryu}, \citenamefont
  {Ko}, \citenamefont {Lee}, \citenamefont {Lee}, \citenamefont {Yoo},
  \citenamefont {Lee}, \citenamefont {Shin}, \citenamefont {Kim},\ and\
  \citenamefont {Byun}}]{Kwon2024NatElecMoS2wafer}%
  \BibitemOpen
  \bibfield  {author} {\bibinfo {author} {\bibfnamefont {J.}~\bibnamefont
  {Kwon}}, \bibinfo {author} {\bibfnamefont {M.}~\bibnamefont {Seol}}, \bibinfo
  {author} {\bibfnamefont {J.}~\bibnamefont {Yoo}}, \bibinfo {author}
  {\bibfnamefont {H.}~\bibnamefont {Ryu}}, \bibinfo {author} {\bibfnamefont
  {D.~S.}\ \bibnamefont {Ko}}, \bibinfo {author} {\bibfnamefont {M.~H.}\
  \bibnamefont {Lee}}, \bibinfo {author} {\bibfnamefont {E.~K.}\ \bibnamefont
  {Lee}}, \bibinfo {author} {\bibfnamefont {M.~S.}\ \bibnamefont {Yoo}},
  \bibinfo {author} {\bibfnamefont {G.~H.}\ \bibnamefont {Lee}}, \bibinfo
  {author} {\bibfnamefont {H.~J.}\ \bibnamefont {Shin}}, \bibinfo {author}
  {\bibfnamefont {J.}~\bibnamefont {Kim}},\ and\ \bibinfo {author}
  {\bibfnamefont {K.~E.}\ \bibnamefont {Byun}},\ }\bibfield  {title} {\bibinfo
  {title} {200-mm-wafer-scale integration of polycrystalline molybdenum
  disulfide transistors},\ }\href {https://doi.org/10.1038/s41928-024-01158-4}
  {\bibfield  {journal} {\bibinfo  {journal} {Nature Electronics}\ }\textbf
  {\bibinfo {volume} {7}},\ \bibinfo {pages} {356} (\bibinfo {year}
  {2024})}\BibitemShut {NoStop}%
\bibitem [{\citenamefont {Kang}\ \emph {et~al.}(2015)\citenamefont {Kang},
  \citenamefont {Xie}, \citenamefont {Huang}, \citenamefont {Han},
  \citenamefont {Huang}, \citenamefont {Mak}, \citenamefont {Kim},
  \citenamefont {Muller},\ and\ \citenamefont
  {Park}}]{Kang2015NatureHighMobility}%
  \BibitemOpen
  \bibfield  {author} {\bibinfo {author} {\bibfnamefont {K.}~\bibnamefont
  {Kang}}, \bibinfo {author} {\bibfnamefont {S.}~\bibnamefont {Xie}}, \bibinfo
  {author} {\bibfnamefont {L.}~\bibnamefont {Huang}}, \bibinfo {author}
  {\bibfnamefont {Y.}~\bibnamefont {Han}}, \bibinfo {author} {\bibfnamefont
  {P.~Y.}\ \bibnamefont {Huang}}, \bibinfo {author} {\bibfnamefont {K.~F.}\
  \bibnamefont {Mak}}, \bibinfo {author} {\bibfnamefont {C.-J.}\ \bibnamefont
  {Kim}}, \bibinfo {author} {\bibfnamefont {D.~A.}\ \bibnamefont {Muller}},\
  and\ \bibinfo {author} {\bibfnamefont {J.}~\bibnamefont {Park}},\ }\bibfield
  {title} {\bibinfo {title} {High-mobility three-atom-thick semiconducting
  films with wafer-scale homogeneity},\ }\href
  {https://doi.org/10.1038/nature14417} {\bibfield  {journal} {\bibinfo
  {journal} {Nature}\ }\textbf {\bibinfo {volume} {520}},\ \bibinfo {pages}
  {656} (\bibinfo {year} {2015})}\BibitemShut {NoStop}%
\bibitem [{\citenamefont {Aspiotis}\ \emph {et~al.}(2023)\citenamefont
  {Aspiotis}, \citenamefont {Morgan}, \citenamefont {M{\"a}rz}, \citenamefont
  {M{\"u}ller-Caspary}, \citenamefont {Ebert}, \citenamefont {Weatherby},
  \citenamefont {Light}, \citenamefont {Huang}, \citenamefont {Hewak},
  \citenamefont {Majumdar},\ and\ \citenamefont {Zeimpekis}}]{Aspiotis2023}%
  \BibitemOpen
  \bibfield  {author} {\bibinfo {author} {\bibfnamefont {N.}~\bibnamefont
  {Aspiotis}}, \bibinfo {author} {\bibfnamefont {K.}~\bibnamefont {Morgan}},
  \bibinfo {author} {\bibfnamefont {B.}~\bibnamefont {M{\"a}rz}}, \bibinfo
  {author} {\bibfnamefont {K.}~\bibnamefont {M{\"u}ller-Caspary}}, \bibinfo
  {author} {\bibfnamefont {M.}~\bibnamefont {Ebert}}, \bibinfo {author}
  {\bibfnamefont {E.}~\bibnamefont {Weatherby}}, \bibinfo {author}
  {\bibfnamefont {M.~E.}\ \bibnamefont {Light}}, \bibinfo {author}
  {\bibfnamefont {C.-C.}\ \bibnamefont {Huang}}, \bibinfo {author}
  {\bibfnamefont {D.~W.}\ \bibnamefont {Hewak}}, \bibinfo {author}
  {\bibfnamefont {S.}~\bibnamefont {Majumdar}},\ and\ \bibinfo {author}
  {\bibfnamefont {I.}~\bibnamefont {Zeimpekis}},\ }\bibfield  {title} {\bibinfo
  {title} {Large-area synthesis of high electrical performance mos2 by a
  commercially scalable atomic layer deposition process},\ }\href
  {https://doi.org/10.1038/s41699-023-00379-z} {\bibfield  {journal} {\bibinfo
  {journal} {npj 2D Materials and Applications}\ }\textbf {\bibinfo {volume}
  {7}},\ \bibinfo {pages} {18} (\bibinfo {year} {2023})}\BibitemShut {NoStop}%
\bibitem [{\citenamefont {Haastrup}\ \emph {et~al.}(2018)\citenamefont
  {Haastrup}, \citenamefont {Strange}, \citenamefont {Pandey}, \citenamefont
  {Deilmann}, \citenamefont {Schmidt}, \citenamefont {Hinsche}, \citenamefont
  {Gjerding}, \citenamefont {Torelli}, \citenamefont {Larsen}, \citenamefont
  {Riis-Jensen}, \citenamefont {Gath}, \citenamefont {Jacobsen}, \citenamefont
  {Mortensen}, \citenamefont {Olsen},\ and\ \citenamefont
  {Thygesen}}]{Haastrup2018_C2DB}%
  \BibitemOpen
  \bibfield  {author} {\bibinfo {author} {\bibfnamefont {S.}~\bibnamefont
  {Haastrup}}, \bibinfo {author} {\bibfnamefont {M.}~\bibnamefont {Strange}},
  \bibinfo {author} {\bibfnamefont {M.}~\bibnamefont {Pandey}}, \bibinfo
  {author} {\bibfnamefont {T.}~\bibnamefont {Deilmann}}, \bibinfo {author}
  {\bibfnamefont {P.~S.}\ \bibnamefont {Schmidt}}, \bibinfo {author}
  {\bibfnamefont {N.~F.}\ \bibnamefont {Hinsche}}, \bibinfo {author}
  {\bibfnamefont {M.~N.}\ \bibnamefont {Gjerding}}, \bibinfo {author}
  {\bibfnamefont {D.}~\bibnamefont {Torelli}}, \bibinfo {author} {\bibfnamefont
  {P.~M.}\ \bibnamefont {Larsen}}, \bibinfo {author} {\bibfnamefont {A.~C.}\
  \bibnamefont {Riis-Jensen}}, \bibinfo {author} {\bibfnamefont
  {J.}~\bibnamefont {Gath}}, \bibinfo {author} {\bibfnamefont {K.~W.}\
  \bibnamefont {Jacobsen}}, \bibinfo {author} {\bibfnamefont {J.~J.}\
  \bibnamefont {Mortensen}}, \bibinfo {author} {\bibfnamefont {T.}~\bibnamefont
  {Olsen}},\ and\ \bibinfo {author} {\bibfnamefont {K.~S.}\ \bibnamefont
  {Thygesen}},\ }\bibfield  {title} {\bibinfo {title} {The computational 2d
  materials database: High-throughput modeling and discovery of atomically thin
  crystals},\ }\href {https://doi.org/10.1088/2053-1583/aacfc1} {\bibfield
  {journal} {\bibinfo  {journal} {2D Materials}\ }\textbf {\bibinfo {volume}
  {5}},\ \bibinfo {pages} {042002} (\bibinfo {year} {2018})}\BibitemShut
  {NoStop}%
\bibitem [{\citenamefont {Zhou}\ \emph {et~al.}(2019)\citenamefont {Zhou},
  \citenamefont {Shen}, \citenamefont {Costa}, \citenamefont {Persson},\ and\
  \citenamefont {Ong}}]{Zhou2019}%
  \BibitemOpen
  \bibfield  {author} {\bibinfo {author} {\bibfnamefont {J.}~\bibnamefont
  {Zhou}}, \bibinfo {author} {\bibfnamefont {L.}~\bibnamefont {Shen}}, \bibinfo
  {author} {\bibfnamefont {M.~D.}\ \bibnamefont {Costa}}, \bibinfo {author}
  {\bibfnamefont {K.~A.}\ \bibnamefont {Persson}},\ and\ \bibinfo {author}
  {\bibfnamefont {S.~P.}\ \bibnamefont {Ong}},\ }\bibfield  {title} {\bibinfo
  {title} {2dmatpedia, an open computational database of two-dimensional
  materials from top-down and bottom-up approaches},\ }\href
  {https://doi.org/10.1038/s41597-019-0097-3} {\bibfield  {journal} {\bibinfo
  {journal} {Scientific Data}\ }\textbf {\bibinfo {volume} {6}},\ \bibinfo
  {pages} {86} (\bibinfo {year} {2019})}\BibitemShut {NoStop}%
\bibitem [{\citenamefont {Campi}\ \emph {et~al.}(2023)\citenamefont {Campi},
  \citenamefont {Mounet}, \citenamefont {Gibertini}, \citenamefont {Pizzi},\
  and\ \citenamefont {Marzari}}]{Campi2023}%
  \BibitemOpen
  \bibfield  {author} {\bibinfo {author} {\bibfnamefont {D.}~\bibnamefont
  {Campi}}, \bibinfo {author} {\bibfnamefont {N.}~\bibnamefont {Mounet}},
  \bibinfo {author} {\bibfnamefont {M.}~\bibnamefont {Gibertini}}, \bibinfo
  {author} {\bibfnamefont {G.}~\bibnamefont {Pizzi}},\ and\ \bibinfo {author}
  {\bibfnamefont {N.}~\bibnamefont {Marzari}},\ }\bibfield  {title} {\bibinfo
  {title} {Expansion of the materials cloud 2d database},\ }\href
  {https://doi.org/10.1021/acsnano.2c11510} {\bibfield  {journal} {\bibinfo
  {journal} {ACS Nano}\ }\textbf {\bibinfo {volume} {17}},\ \bibinfo {pages}
  {11268} (\bibinfo {year} {2023})}\BibitemShut {NoStop}%
\bibitem [{\citenamefont {Mounet}\ \emph {et~al.}(2018)\citenamefont {Mounet},
  \citenamefont {Gibertini}, \citenamefont {Schwaller}, \citenamefont {Campi},
  \citenamefont {Merkys}, \citenamefont {Marrazzo}, \citenamefont {Sohier},
  \citenamefont {Castelli}, \citenamefont {Cepellotti}, \citenamefont {Pizzi},\
  and\ \citenamefont {Marzari}}]{Mounet2018}%
  \BibitemOpen
  \bibfield  {author} {\bibinfo {author} {\bibfnamefont {N.}~\bibnamefont
  {Mounet}}, \bibinfo {author} {\bibfnamefont {M.}~\bibnamefont {Gibertini}},
  \bibinfo {author} {\bibfnamefont {P.}~\bibnamefont {Schwaller}}, \bibinfo
  {author} {\bibfnamefont {D.}~\bibnamefont {Campi}}, \bibinfo {author}
  {\bibfnamefont {A.}~\bibnamefont {Merkys}}, \bibinfo {author} {\bibfnamefont
  {A.}~\bibnamefont {Marrazzo}}, \bibinfo {author} {\bibfnamefont
  {T.}~\bibnamefont {Sohier}}, \bibinfo {author} {\bibfnamefont {I.~E.}\
  \bibnamefont {Castelli}}, \bibinfo {author} {\bibfnamefont {A.}~\bibnamefont
  {Cepellotti}}, \bibinfo {author} {\bibfnamefont {G.}~\bibnamefont {Pizzi}},\
  and\ \bibinfo {author} {\bibfnamefont {N.}~\bibnamefont {Marzari}},\
  }\bibfield  {title} {\bibinfo {title} {Two-dimensional materials from
  high-throughput computational exfoliation of experimentally known
  compounds},\ }\href {https://doi.org/10.1038/s41565-017-0035-5} {\bibfield
  {journal} {\bibinfo  {journal} {Nature Nanotechnology}\ }\textbf {\bibinfo
  {volume} {13}},\ \bibinfo {pages} {246} (\bibinfo {year} {2018})}\BibitemShut
  {NoStop}%
\bibitem [{\citenamefont {Choudhary}\ \emph {et~al.}(2020)\citenamefont
  {Choudhary}, \citenamefont {Garrity}, \citenamefont {Reid}, \citenamefont
  {DeCost}, \citenamefont {Biacchi}, \citenamefont {Hight~Walker},
  \citenamefont {Trautt}, \citenamefont {Hattrick-Simpers}, \citenamefont
  {Kusne}, \citenamefont {Centrone}, \citenamefont {Davydov}, \citenamefont
  {Jiang}, \citenamefont {Pachter}, \citenamefont {Cheon}, \citenamefont
  {Reed}, \citenamefont {Agrawal}, \citenamefont {Qian}, \citenamefont
  {Sharma}, \citenamefont {Kalinin}, \citenamefont {Sumpter}, \citenamefont
  {Pilania}, \citenamefont {Acar}, \citenamefont {Mandal}, \citenamefont
  {Haule}, \citenamefont {Vanderbilt}, \citenamefont {Rabe},\ and\
  \citenamefont {Tavazza}}]{Choudhary2020_JARVIS}%
  \BibitemOpen
  \bibfield  {author} {\bibinfo {author} {\bibfnamefont {K.}~\bibnamefont
  {Choudhary}}, \bibinfo {author} {\bibfnamefont {K.~F.}\ \bibnamefont
  {Garrity}}, \bibinfo {author} {\bibfnamefont {A.~C.~E.}\ \bibnamefont
  {Reid}}, \bibinfo {author} {\bibfnamefont {B.}~\bibnamefont {DeCost}},
  \bibinfo {author} {\bibfnamefont {A.~J.}\ \bibnamefont {Biacchi}}, \bibinfo
  {author} {\bibfnamefont {A.~R.}\ \bibnamefont {Hight~Walker}}, \bibinfo
  {author} {\bibfnamefont {Z.}~\bibnamefont {Trautt}}, \bibinfo {author}
  {\bibfnamefont {J.}~\bibnamefont {Hattrick-Simpers}}, \bibinfo {author}
  {\bibfnamefont {A.~G.}\ \bibnamefont {Kusne}}, \bibinfo {author}
  {\bibfnamefont {A.}~\bibnamefont {Centrone}}, \bibinfo {author}
  {\bibfnamefont {A.}~\bibnamefont {Davydov}}, \bibinfo {author} {\bibfnamefont
  {J.}~\bibnamefont {Jiang}}, \bibinfo {author} {\bibfnamefont
  {R.}~\bibnamefont {Pachter}}, \bibinfo {author} {\bibfnamefont
  {G.}~\bibnamefont {Cheon}}, \bibinfo {author} {\bibfnamefont
  {E.}~\bibnamefont {Reed}}, \bibinfo {author} {\bibfnamefont {A.}~\bibnamefont
  {Agrawal}}, \bibinfo {author} {\bibfnamefont {X.}~\bibnamefont {Qian}},
  \bibinfo {author} {\bibfnamefont {V.}~\bibnamefont {Sharma}}, \bibinfo
  {author} {\bibfnamefont {S.~V.}\ \bibnamefont {Kalinin}}, \bibinfo {author}
  {\bibfnamefont {B.~G.}\ \bibnamefont {Sumpter}}, \bibinfo {author}
  {\bibfnamefont {G.}~\bibnamefont {Pilania}}, \bibinfo {author} {\bibfnamefont
  {P.}~\bibnamefont {Acar}}, \bibinfo {author} {\bibfnamefont {S.}~\bibnamefont
  {Mandal}}, \bibinfo {author} {\bibfnamefont {K.}~\bibnamefont {Haule}},
  \bibinfo {author} {\bibfnamefont {D.}~\bibnamefont {Vanderbilt}}, \bibinfo
  {author} {\bibfnamefont {K.}~\bibnamefont {Rabe}},\ and\ \bibinfo {author}
  {\bibfnamefont {F.}~\bibnamefont {Tavazza}},\ }\bibfield  {title} {\bibinfo
  {title} {The joint automated repository for various integrated simulations
  (jarvis) for data-driven materials design},\ }\href
  {https://doi.org/10.1038/s41524-020-00440-1} {\bibfield  {journal} {\bibinfo
  {journal} {npj Computational Materials}\ }\textbf {\bibinfo {volume} {6}},\
  \bibinfo {pages} {173} (\bibinfo {year} {2020})}\BibitemShut {NoStop}%
\bibitem [{\citenamefont {Shen}\ \emph {et~al.}(2022)\citenamefont {Shen},
  \citenamefont {Zhou}, \citenamefont {Yang}, \citenamefont {Yang},
  \citenamefont {Feng}, \citenamefont {Zhang}, \citenamefont {Wang},
  \citenamefont {Liu},\ and\ \citenamefont {Wang}}]{Shen2022_2DHT_Discovery}%
  \BibitemOpen
  \bibfield  {author} {\bibinfo {author} {\bibfnamefont {L.}~\bibnamefont
  {Shen}}, \bibinfo {author} {\bibfnamefont {J.}~\bibnamefont {Zhou}}, \bibinfo
  {author} {\bibfnamefont {T.}~\bibnamefont {Yang}}, \bibinfo {author}
  {\bibfnamefont {M.}~\bibnamefont {Yang}}, \bibinfo {author} {\bibfnamefont
  {Y.~P.}\ \bibnamefont {Feng}}, \bibinfo {author} {\bibfnamefont
  {Y.}~\bibnamefont {Zhang}}, \bibinfo {author} {\bibfnamefont
  {Q.}~\bibnamefont {Wang}}, \bibinfo {author} {\bibfnamefont {J.}~\bibnamefont
  {Liu}},\ and\ \bibinfo {author} {\bibfnamefont {J.}~\bibnamefont {Wang}},\
  }\bibfield  {title} {\bibinfo {title} {High-throughput computational
  discovery and intelligent design of two-dimensional functional materials for
  various applications},\ }\href {https://doi.org/10.1021/accountsmr.1c00246}
  {\bibfield  {journal} {\bibinfo  {journal} {Accounts of Materials Research}\
  }\textbf {\bibinfo {volume} {3}},\ \bibinfo {pages} {572} (\bibinfo {year}
  {2022})}\BibitemShut {NoStop}%
\bibitem [{\citenamefont {Lu}\ \emph {et~al.}(2024)\citenamefont {Lu},
  \citenamefont {Xia}, \citenamefont {Ren}, \citenamefont {Xie}, \citenamefont
  {Zhou}, \citenamefont {Vinai}, \citenamefont {Morton}, \citenamefont {Wee},
  \citenamefont {van~der Wiel}, \citenamefont {Zhang},\ and\ \citenamefont
  {Wong}}]{Lu2024_ML_2D_Review}%
  \BibitemOpen
  \bibfield  {author} {\bibinfo {author} {\bibfnamefont {B.}~\bibnamefont
  {Lu}}, \bibinfo {author} {\bibfnamefont {Y.}~\bibnamefont {Xia}}, \bibinfo
  {author} {\bibfnamefont {Y.}~\bibnamefont {Ren}}, \bibinfo {author}
  {\bibfnamefont {M.}~\bibnamefont {Xie}}, \bibinfo {author} {\bibfnamefont
  {L.}~\bibnamefont {Zhou}}, \bibinfo {author} {\bibfnamefont {G.}~\bibnamefont
  {Vinai}}, \bibinfo {author} {\bibfnamefont {S.~A.}\ \bibnamefont {Morton}},
  \bibinfo {author} {\bibfnamefont {A.~T.~S.}\ \bibnamefont {Wee}}, \bibinfo
  {author} {\bibfnamefont {W.~G.}\ \bibnamefont {van~der Wiel}}, \bibinfo
  {author} {\bibfnamefont {W.}~\bibnamefont {Zhang}},\ and\ \bibinfo {author}
  {\bibfnamefont {P.~K.~J.}\ \bibnamefont {Wong}},\ }\bibfield  {title}
  {\bibinfo {title} {When machine learning meets 2d materials: A review},\
  }\href {https://doi.org/10.1002/advs.202305277} {\bibfield  {journal}
  {\bibinfo  {journal} {Advanced Science}\ }\textbf {\bibinfo {volume} {11}},\
  \bibinfo {pages} {2305277} (\bibinfo {year} {2024})}\BibitemShut {NoStop}%
\bibitem [{\citenamefont {Kormányos}\ \emph {et~al.}(2015)\citenamefont
  {Kormányos}, \citenamefont {Burkard}, \citenamefont {Gmitra}, \citenamefont
  {Fabian}, \citenamefont {Zólyomi}, \citenamefont {Drummond},\ and\
  \citenamefont {Fal’ko}}]{Kormányos_2015}%
  \BibitemOpen
  \bibfield  {author} {\bibinfo {author} {\bibfnamefont {A.}~\bibnamefont
  {Kormányos}}, \bibinfo {author} {\bibfnamefont {G.}~\bibnamefont {Burkard}},
  \bibinfo {author} {\bibfnamefont {M.}~\bibnamefont {Gmitra}}, \bibinfo
  {author} {\bibfnamefont {J.}~\bibnamefont {Fabian}}, \bibinfo {author}
  {\bibfnamefont {V.}~\bibnamefont {Zólyomi}}, \bibinfo {author}
  {\bibfnamefont {N.~D.}\ \bibnamefont {Drummond}},\ and\ \bibinfo {author}
  {\bibfnamefont {V.}~\bibnamefont {Fal’ko}},\ }\bibfield  {title} {\bibinfo
  {title} {k·p theory for two-dimensional transition metal dichalcogenide
  semiconductors},\ }\href {https://doi.org/10.1088/2053-1583/2/2/022001}
  {\bibfield  {journal} {\bibinfo  {journal} {2D Materials}\ }\textbf {\bibinfo
  {volume} {2}},\ \bibinfo {pages} {022001} (\bibinfo {year}
  {2015})}\BibitemShut {NoStop}%
\bibitem [{\citenamefont {Korm{\'a}nyos}\ \emph {et~al.}(2013)\citenamefont
  {Korm{\'a}nyos}, \citenamefont {Z{\'o}lyomi}, \citenamefont {Drummond},
  \citenamefont {Rakyta}, \citenamefont {Burkard},\ and\ \citenamefont
  {Fal'ko}}]{Kormanyos2013_MoS2_TW}%
  \BibitemOpen
  \bibfield  {author} {\bibinfo {author} {\bibfnamefont {A.}~\bibnamefont
  {Korm{\'a}nyos}}, \bibinfo {author} {\bibfnamefont {V.}~\bibnamefont
  {Z{\'o}lyomi}}, \bibinfo {author} {\bibfnamefont {N.~D.}\ \bibnamefont
  {Drummond}}, \bibinfo {author} {\bibfnamefont {P.}~\bibnamefont {Rakyta}},
  \bibinfo {author} {\bibfnamefont {G.}~\bibnamefont {Burkard}},\ and\ \bibinfo
  {author} {\bibfnamefont {V.~I.}\ \bibnamefont {Fal'ko}},\ }\bibfield  {title}
  {\bibinfo {title} {Monolayer mos$_2$: Trigonal warping, the $\gamma$ valley,
  and spin-orbit coupling effects},\ }\href
  {https://doi.org/10.1103/PhysRevB.88.045416} {\bibfield  {journal} {\bibinfo
  {journal} {Physical Review B}\ }\textbf {\bibinfo {volume} {88}},\ \bibinfo
  {pages} {045416} (\bibinfo {year} {2013})}\BibitemShut {NoStop}%
\bibitem [{\citenamefont {Ko\ifmmode~\acute{s}\else \'{s}\fi{}mider}\ \emph
  {et~al.}(2013)\citenamefont {Ko\ifmmode~\acute{s}\else \'{s}\fi{}mider},
  \citenamefont {Gonz\'alez},\ and\ \citenamefont
  {Fern\'andez-Rossier}}]{PhysRevB.88.245436}%
  \BibitemOpen
  \bibfield  {author} {\bibinfo {author} {\bibfnamefont {K.}~\bibnamefont
  {Ko\ifmmode~\acute{s}\else \'{s}\fi{}mider}}, \bibinfo {author}
  {\bibfnamefont {J.~W.}\ \bibnamefont {Gonz\'alez}},\ and\ \bibinfo {author}
  {\bibfnamefont {J.}~\bibnamefont {Fern\'andez-Rossier}},\ }\bibfield  {title}
  {\bibinfo {title} {Large spin splitting in the conduction band of transition
  metal dichalcogenide monolayers},\ }\href
  {https://doi.org/10.1103/PhysRevB.88.245436} {\bibfield  {journal} {\bibinfo
  {journal} {Phys. Rev. B}\ }\textbf {\bibinfo {volume} {88}},\ \bibinfo
  {pages} {245436} (\bibinfo {year} {2013})}\BibitemShut {NoStop}%
\bibitem [{\citenamefont {Cheiwchanchamnangij}\ \emph
  {et~al.}(2013)\citenamefont {Cheiwchanchamnangij}, \citenamefont {Lambrecht},
  \citenamefont {Song},\ and\ \citenamefont
  {Dery}}]{Cheiwchanchamnangij2013_StrainSOC}%
  \BibitemOpen
  \bibfield  {author} {\bibinfo {author} {\bibfnamefont {T.}~\bibnamefont
  {Cheiwchanchamnangij}}, \bibinfo {author} {\bibfnamefont {W.~R.~L.}\
  \bibnamefont {Lambrecht}}, \bibinfo {author} {\bibfnamefont {Y.}~\bibnamefont
  {Song}},\ and\ \bibinfo {author} {\bibfnamefont {H.}~\bibnamefont {Dery}},\
  }\bibfield  {title} {\bibinfo {title} {Strain effects on the
  spin-orbit-induced band structure splittings in monolayer mos$_2$ and
  graphene},\ }\href {https://doi.org/10.1103/PhysRevB.88.155404} {\bibfield
  {journal} {\bibinfo  {journal} {Physical Review B}\ }\textbf {\bibinfo
  {volume} {88}},\ \bibinfo {pages} {155404} (\bibinfo {year}
  {2013})}\BibitemShut {NoStop}%
\bibitem [{\citenamefont {Rodrigues~Pela}\ \emph {et~al.}(2024)\citenamefont
  {Rodrigues~Pela}, \citenamefont {Vona}, \citenamefont {Lubeck}, \citenamefont
  {Alex}, \citenamefont {Gonzalez~Oliva},\ and\ \citenamefont
  {Draxl}}]{RodriguesPela2024}%
  \BibitemOpen
  \bibfield  {author} {\bibinfo {author} {\bibfnamefont {R.}~\bibnamefont
  {Rodrigues~Pela}}, \bibinfo {author} {\bibfnamefont {C.}~\bibnamefont
  {Vona}}, \bibinfo {author} {\bibfnamefont {S.}~\bibnamefont {Lubeck}},
  \bibinfo {author} {\bibfnamefont {B.}~\bibnamefont {Alex}}, \bibinfo {author}
  {\bibfnamefont {I.}~\bibnamefont {Gonzalez~Oliva}},\ and\ \bibinfo {author}
  {\bibfnamefont {C.}~\bibnamefont {Draxl}},\ }\bibfield  {title} {\bibinfo
  {title} {Critical assessment of g0w0 calculations for 2d materials: the
  example of monolayer mos2},\ }\href
  {https://doi.org/10.1038/s41524-024-01253-2} {\bibfield  {journal} {\bibinfo
  {journal} {npj Computational Materials}\ }\textbf {\bibinfo {volume} {10}},\
  \bibinfo {pages} {77} (\bibinfo {year} {2024})}\BibitemShut {NoStop}%
\bibitem [{\citenamefont {Marinov}\ \emph {et~al.}(2017)\citenamefont
  {Marinov}, \citenamefont {Avsar}, \citenamefont {Watanabe}, \citenamefont
  {Taniguchi},\ and\ \citenamefont {Kis}}]{Marinov2017}%
  \BibitemOpen
  \bibfield  {author} {\bibinfo {author} {\bibfnamefont {K.}~\bibnamefont
  {Marinov}}, \bibinfo {author} {\bibfnamefont {A.}~\bibnamefont {Avsar}},
  \bibinfo {author} {\bibfnamefont {K.}~\bibnamefont {Watanabe}}, \bibinfo
  {author} {\bibfnamefont {T.}~\bibnamefont {Taniguchi}},\ and\ \bibinfo
  {author} {\bibfnamefont {A.}~\bibnamefont {Kis}},\ }\bibfield  {title}
  {\bibinfo {title} {{Resolving the spin splitting in the conduction band of
  monolayer MoS2}},\ }\href {https://doi.org/10.1038/s41467-017-02047-5}
  {\bibfield  {journal} {\bibinfo  {journal} {Nature Communications}\ }\textbf
  {\bibinfo {volume} {8}},\ \bibinfo {pages} {1938} (\bibinfo {year}
  {2017})}\BibitemShut {NoStop}%
\bibitem [{\citenamefont {Pisoni}\ \emph {et~al.}(2018)\citenamefont {Pisoni},
  \citenamefont {Korm\'anyos}, \citenamefont {Brooks}, \citenamefont {Lei},
  \citenamefont {Back}, \citenamefont {Eich}, \citenamefont {Overweg},
  \citenamefont {Lee}, \citenamefont {Rickhaus}, \citenamefont {Watanabe},
  \citenamefont {Taniguchi}, \citenamefont {Imamoglu}, \citenamefont {Burkard},
  \citenamefont {Ihn},\ and\ \citenamefont {Ensslin}}]{PhysRevLett.121.247701}%
  \BibitemOpen
  \bibfield  {author} {\bibinfo {author} {\bibfnamefont {R.}~\bibnamefont
  {Pisoni}}, \bibinfo {author} {\bibfnamefont {A.}~\bibnamefont {Korm\'anyos}},
  \bibinfo {author} {\bibfnamefont {M.}~\bibnamefont {Brooks}}, \bibinfo
  {author} {\bibfnamefont {Z.}~\bibnamefont {Lei}}, \bibinfo {author}
  {\bibfnamefont {P.}~\bibnamefont {Back}}, \bibinfo {author} {\bibfnamefont
  {M.}~\bibnamefont {Eich}}, \bibinfo {author} {\bibfnamefont {H.}~\bibnamefont
  {Overweg}}, \bibinfo {author} {\bibfnamefont {Y.}~\bibnamefont {Lee}},
  \bibinfo {author} {\bibfnamefont {P.}~\bibnamefont {Rickhaus}}, \bibinfo
  {author} {\bibfnamefont {K.}~\bibnamefont {Watanabe}}, \bibinfo {author}
  {\bibfnamefont {T.}~\bibnamefont {Taniguchi}}, \bibinfo {author}
  {\bibfnamefont {A.}~\bibnamefont {Imamoglu}}, \bibinfo {author}
  {\bibfnamefont {G.}~\bibnamefont {Burkard}}, \bibinfo {author} {\bibfnamefont
  {T.}~\bibnamefont {Ihn}},\ and\ \bibinfo {author} {\bibfnamefont
  {K.}~\bibnamefont {Ensslin}},\ }\bibfield  {title} {\bibinfo {title}
  {{Interactions and Magnetotransport through Spin-Valley Coupled Landau Levels
  in Monolayer ${\mathrm{MoS}}_{2}$}},\ }\href
  {https://doi.org/10.1103/PhysRevLett.121.247701} {\bibfield  {journal}
  {\bibinfo  {journal} {Phys. Rev. Lett.}\ }\textbf {\bibinfo {volume} {121}},\
  \bibinfo {pages} {247701} (\bibinfo {year} {2018})}\BibitemShut {NoStop}%
\bibitem [{\citenamefont {Lin}\ \emph {et~al.}(2019)\citenamefont {Lin},
  \citenamefont {Han}, \citenamefont {Piot}, \citenamefont {Wu}, \citenamefont
  {Xu}, \citenamefont {Long}, \citenamefont {An}, \citenamefont {Cheung},
  \citenamefont {Zheng}, \citenamefont {Plochocka}, \citenamefont {Dai},
  \citenamefont {Maude}, \citenamefont {Zhang},\ and\ \citenamefont
  {Wang}}]{Lin2019_MoS2_ValleySusceptibility}%
  \BibitemOpen
  \bibfield  {author} {\bibinfo {author} {\bibfnamefont {J.}~\bibnamefont
  {Lin}}, \bibinfo {author} {\bibfnamefont {T.}~\bibnamefont {Han}}, \bibinfo
  {author} {\bibfnamefont {B.~A.}\ \bibnamefont {Piot}}, \bibinfo {author}
  {\bibfnamefont {Z.}~\bibnamefont {Wu}}, \bibinfo {author} {\bibfnamefont
  {S.}~\bibnamefont {Xu}}, \bibinfo {author} {\bibfnamefont {G.}~\bibnamefont
  {Long}}, \bibinfo {author} {\bibfnamefont {L.}~\bibnamefont {An}}, \bibinfo
  {author} {\bibfnamefont {P.}~\bibnamefont {Cheung}}, \bibinfo {author}
  {\bibfnamefont {P.-P.}\ \bibnamefont {Zheng}}, \bibinfo {author}
  {\bibfnamefont {P.}~\bibnamefont {Plochocka}}, \bibinfo {author}
  {\bibfnamefont {X.}~\bibnamefont {Dai}}, \bibinfo {author} {\bibfnamefont
  {D.~K.}\ \bibnamefont {Maude}}, \bibinfo {author} {\bibfnamefont
  {F.}~\bibnamefont {Zhang}},\ and\ \bibinfo {author} {\bibfnamefont
  {N.}~\bibnamefont {Wang}},\ }\bibfield  {title} {\bibinfo {title}
  {Determining interaction enhanced valley susceptibility in spin-valley-locked
  mos$_2$},\ }\href {https://doi.org/10.1021/acs.nanolett.8b04731} {\bibfield
  {journal} {\bibinfo  {journal} {Nano Letters}\ }\textbf {\bibinfo {volume}
  {19}},\ \bibinfo {pages} {1736} (\bibinfo {year} {2019})}\BibitemShut
  {NoStop}%
\bibitem [{\citenamefont {Chi}\ \emph {et~al.}(2023)\citenamefont {Chi},
  \citenamefont {Wei}, \citenamefont {Zhang}, \citenamefont {Chen},\ and\
  \citenamefont {Weng}}]{doi:10.1021/acs.jpclett.3c02431}%
  \BibitemOpen
  \bibfield  {author} {\bibinfo {author} {\bibfnamefont {Z.}~\bibnamefont
  {Chi}}, \bibinfo {author} {\bibfnamefont {Z.}~\bibnamefont {Wei}}, \bibinfo
  {author} {\bibfnamefont {G.}~\bibnamefont {Zhang}}, \bibinfo {author}
  {\bibfnamefont {H.}~\bibnamefont {Chen}},\ and\ \bibinfo {author}
  {\bibfnamefont {Y.-X.}\ \bibnamefont {Weng}},\ }\bibfield  {title} {\bibinfo
  {title} {{Determining Band Splitting and Spin-Flip Dynamics in Monolayer
  MoS2}},\ }\href {https://doi.org/10.1021/acs.jpclett.3c02431} {\bibfield
  {journal} {\bibinfo  {journal} {The Journal of Physical Chemistry Letters}\
  }\textbf {\bibinfo {volume} {14}},\ \bibinfo {pages} {9640} (\bibinfo {year}
  {2023})},\ \bibinfo {note} {pMID: 37870497},\ \Eprint
  {https://arxiv.org/abs/https://doi.org/10.1021/acs.jpclett.3c02431}
  {https://doi.org/10.1021/acs.jpclett.3c02431} \BibitemShut {NoStop}%
\bibitem [{\citenamefont {Robert}\ \emph {et~al.}(2020)\citenamefont {Robert},
  \citenamefont {Han}, \citenamefont {Kapuscinski}, \citenamefont {Delhomme},
  \citenamefont {Faugeras}, \citenamefont {Amand}, \citenamefont {Molas},
  \citenamefont {Bartos}, \citenamefont {Watanabe}, \citenamefont {Taniguchi},
  \citenamefont {Urbaszek}, \citenamefont {Potemski},\ and\ \citenamefont
  {Marie}}]{Robert2020}%
  \BibitemOpen
  \bibfield  {author} {\bibinfo {author} {\bibfnamefont {C.}~\bibnamefont
  {Robert}}, \bibinfo {author} {\bibfnamefont {B.}~\bibnamefont {Han}},
  \bibinfo {author} {\bibfnamefont {P.}~\bibnamefont {Kapuscinski}}, \bibinfo
  {author} {\bibfnamefont {A.}~\bibnamefont {Delhomme}}, \bibinfo {author}
  {\bibfnamefont {C.}~\bibnamefont {Faugeras}}, \bibinfo {author}
  {\bibfnamefont {T.}~\bibnamefont {Amand}}, \bibinfo {author} {\bibfnamefont
  {M.~R.}\ \bibnamefont {Molas}}, \bibinfo {author} {\bibfnamefont
  {M.}~\bibnamefont {Bartos}}, \bibinfo {author} {\bibfnamefont
  {K.}~\bibnamefont {Watanabe}}, \bibinfo {author} {\bibfnamefont
  {T.}~\bibnamefont {Taniguchi}}, \bibinfo {author} {\bibfnamefont
  {B.}~\bibnamefont {Urbaszek}}, \bibinfo {author} {\bibfnamefont
  {M.}~\bibnamefont {Potemski}},\ and\ \bibinfo {author} {\bibfnamefont
  {X.}~\bibnamefont {Marie}},\ }\bibfield  {title} {\bibinfo {title}
  {{Measurement of the spin-forbidden dark excitons in MoS2 and MoSe2
  monolayers}},\ }\href {https://doi.org/10.1038/s41467-020-17608-4} {\bibfield
   {journal} {\bibinfo  {journal} {Nature Communications}\ }\textbf {\bibinfo
  {volume} {11}},\ \bibinfo {pages} {4037} (\bibinfo {year}
  {2020})}\BibitemShut {NoStop}%
\bibitem [{\citenamefont {Rozhansky}\ and\ \citenamefont
  {Fal'ko}(2024)}]{PhysRevB.110.L161404}%
  \BibitemOpen
  \bibfield  {author} {\bibinfo {author} {\bibfnamefont {I.}~\bibnamefont
  {Rozhansky}}\ and\ \bibinfo {author} {\bibfnamefont {V.}~\bibnamefont
  {Fal'ko}},\ }\bibfield  {title} {\bibinfo {title} {Exchange-enhanced
  spin-orbit splitting and its density dependence for electrons in monolayer
  transition metal dichalcogenides},\ }\href
  {https://doi.org/10.1103/PhysRevB.110.L161404} {\bibfield  {journal}
  {\bibinfo  {journal} {Phys. Rev. B}\ }\textbf {\bibinfo {volume} {110}},\
  \bibinfo {pages} {L161404} (\bibinfo {year} {2024})}\BibitemShut {NoStop}%
\bibitem [{\citenamefont {Scharf}\ \emph {et~al.}(2019)\citenamefont {Scharf},
  \citenamefont {Tuan}, \citenamefont {Žutić},\ and\ \citenamefont
  {Dery}}]{Scharf_2019}%
  \BibitemOpen
  \bibfield  {author} {\bibinfo {author} {\bibfnamefont {B.}~\bibnamefont
  {Scharf}}, \bibinfo {author} {\bibfnamefont {D.~V.}\ \bibnamefont {Tuan}},
  \bibinfo {author} {\bibfnamefont {I.}~\bibnamefont {Žutić}},\ and\ \bibinfo
  {author} {\bibfnamefont {H.}~\bibnamefont {Dery}},\ }\bibfield  {title}
  {\bibinfo {title} {Dynamical screening in monolayer transition-metal
  dichalcogenides and its manifestations in the exciton spectrum},\ }\href
  {https://doi.org/10.1088/1361-648X/ab071f} {\bibfield  {journal} {\bibinfo
  {journal} {Journal of Physics: Condensed Matter}\ }\textbf {\bibinfo {volume}
  {31}},\ \bibinfo {pages} {203001} (\bibinfo {year} {2019})}\BibitemShut
  {NoStop}%
\bibitem [{\citenamefont {Ochoa}\ \emph {et~al.}(2014)\citenamefont {Ochoa},
  \citenamefont {Finocchiaro}, \citenamefont {Guinea},\ and\ \citenamefont
  {Fal'ko}}]{PhysRevB.90.235429}%
  \BibitemOpen
  \bibfield  {author} {\bibinfo {author} {\bibfnamefont {H.}~\bibnamefont
  {Ochoa}}, \bibinfo {author} {\bibfnamefont {F.}~\bibnamefont {Finocchiaro}},
  \bibinfo {author} {\bibfnamefont {F.}~\bibnamefont {Guinea}},\ and\ \bibinfo
  {author} {\bibfnamefont {V.~I.}\ \bibnamefont {Fal'ko}},\ }\bibfield  {title}
  {\bibinfo {title} {Spin-valley relaxation and quantum transport regimes in
  two-dimensional transition-metal dichalcogenides},\ }\href
  {https://doi.org/10.1103/PhysRevB.90.235429} {\bibfield  {journal} {\bibinfo
  {journal} {Phys. Rev. B}\ }\textbf {\bibinfo {volume} {90}},\ \bibinfo
  {pages} {235429} (\bibinfo {year} {2014})}\BibitemShut {NoStop}%
\bibitem [{sup()}]{supp}%
  \BibitemOpen
  \href@noop {} {}\bibinfo {howpublished}
  {\url{URL_will_be_inserted_by_publisher}}\BibitemShut {NoStop}%
\bibitem [{\citenamefont {Eknapakul}\ \emph {et~al.}(2014)\citenamefont
  {Eknapakul}, \citenamefont {King}, \citenamefont {Asakawa}, \citenamefont
  {Buaphet}, \citenamefont {He}, \citenamefont {Mo}, \citenamefont {Takagi},
  \citenamefont {Shen}, \citenamefont {Baumberger}, \citenamefont {Sasagawa},
  \citenamefont {Jungthawan},\ and\ \citenamefont {Meevasana}}]{Eknapakul2014}%
  \BibitemOpen
  \bibfield  {author} {\bibinfo {author} {\bibfnamefont {T.}~\bibnamefont
  {Eknapakul}}, \bibinfo {author} {\bibfnamefont {P.~D.~C.}\ \bibnamefont
  {King}}, \bibinfo {author} {\bibfnamefont {M.}~\bibnamefont {Asakawa}},
  \bibinfo {author} {\bibfnamefont {P.}~\bibnamefont {Buaphet}}, \bibinfo
  {author} {\bibfnamefont {R.-H.}\ \bibnamefont {He}}, \bibinfo {author}
  {\bibfnamefont {S.-K.}\ \bibnamefont {Mo}}, \bibinfo {author} {\bibfnamefont
  {H.}~\bibnamefont {Takagi}}, \bibinfo {author} {\bibfnamefont {K.~M.}\
  \bibnamefont {Shen}}, \bibinfo {author} {\bibfnamefont {F.}~\bibnamefont
  {Baumberger}}, \bibinfo {author} {\bibfnamefont {T.}~\bibnamefont
  {Sasagawa}}, \bibinfo {author} {\bibfnamefont {S.}~\bibnamefont
  {Jungthawan}},\ and\ \bibinfo {author} {\bibfnamefont {W.}~\bibnamefont
  {Meevasana}},\ }\bibfield  {title} {\bibinfo {title} {Electronic structure of
  a quasi-freestanding mos2 monolayer},\ }\href
  {https://doi.org/10.1021/nl4042824} {\bibfield  {journal} {\bibinfo
  {journal} {Nano Letters}\ }\textbf {\bibinfo {volume} {14}},\ \bibinfo
  {pages} {1312} (\bibinfo {year} {2014})}\BibitemShut {NoStop}%
\bibitem [{\citenamefont {Giuliani}\ and\ \citenamefont
  {Vignale}(2005)}]{Giuliani_Vignale_2005}%
  \BibitemOpen
  \bibfield  {author} {\bibinfo {author} {\bibfnamefont {G.}~\bibnamefont
  {Giuliani}}\ and\ \bibinfo {author} {\bibfnamefont {G.}~\bibnamefont
  {Vignale}},\ }\href@noop {} {\emph {\bibinfo {title} {Quantum Theory of the
  Electron Liquid}}}\ (\bibinfo  {publisher} {Cambridge University Press},\
  \bibinfo {year} {2005})\BibitemShut {NoStop}%
\bibitem [{\citenamefont {Das~Sarma}\ \emph {et~al.}(2004)\citenamefont
  {Das~Sarma}, \citenamefont {Galitski},\ and\ \citenamefont
  {Zhang}}]{PhysRevB.69.125334}%
  \BibitemOpen
  \bibfield  {author} {\bibinfo {author} {\bibfnamefont {S.}~\bibnamefont
  {Das~Sarma}}, \bibinfo {author} {\bibfnamefont {V.~M.}\ \bibnamefont
  {Galitski}},\ and\ \bibinfo {author} {\bibfnamefont {Y.}~\bibnamefont
  {Zhang}},\ }\bibfield  {title} {\bibinfo {title} {Temperature-dependent
  effective-mass renormalization in two-dimensional electron systems},\ }\href
  {https://doi.org/10.1103/PhysRevB.69.125334} {\bibfield  {journal} {\bibinfo
  {journal} {Phys. Rev. B}\ }\textbf {\bibinfo {volume} {69}},\ \bibinfo
  {pages} {125334} (\bibinfo {year} {2004})}\BibitemShut {NoStop}%
\bibitem [{\citenamefont {Zhang}\ and\ \citenamefont
  {Das~Sarma}(2004)}]{PhysRevB.70.035104}%
  \BibitemOpen
  \bibfield  {author} {\bibinfo {author} {\bibfnamefont {Y.}~\bibnamefont
  {Zhang}}\ and\ \bibinfo {author} {\bibfnamefont {S.}~\bibnamefont
  {Das~Sarma}},\ }\bibfield  {title} {\bibinfo {title} {Temperature-dependent
  effective mass renormalization in a coulomb fermi liquid},\ }\href
  {https://doi.org/10.1103/PhysRevB.70.035104} {\bibfield  {journal} {\bibinfo
  {journal} {Phys. Rev. B}\ }\textbf {\bibinfo {volume} {70}},\ \bibinfo
  {pages} {035104} (\bibinfo {year} {2004})}\BibitemShut {NoStop}%
\bibitem [{\citenamefont {Galitski}\ and\ \citenamefont
  {Das~Sarma}(2004)}]{PhysRevB.70.035111}%
  \BibitemOpen
  \bibfield  {author} {\bibinfo {author} {\bibfnamefont {V.~M.}\ \bibnamefont
  {Galitski}}\ and\ \bibinfo {author} {\bibfnamefont {S.}~\bibnamefont
  {Das~Sarma}},\ }\bibfield  {title} {\bibinfo {title} {Universal temperature
  corrections to fermi liquid theory in an interacting electron system},\
  }\href {https://doi.org/10.1103/PhysRevB.70.035111} {\bibfield  {journal}
  {\bibinfo  {journal} {Phys. Rev. B}\ }\textbf {\bibinfo {volume} {70}},\
  \bibinfo {pages} {035111} (\bibinfo {year} {2004})}\BibitemShut {NoStop}%
\bibitem [{\citenamefont {Galiautdinov}(2019)}]{GALIAUTDINOV20193167}%
  \BibitemOpen
  \bibfield  {author} {\bibinfo {author} {\bibfnamefont {A.}~\bibnamefont
  {Galiautdinov}},\ }\bibfield  {title} {\bibinfo {title} {Anisotropic keldysh
  interaction},\ }\href
  {https://doi.org/https://doi.org/10.1016/j.physleta.2019.07.002} {\bibfield
  {journal} {\bibinfo  {journal} {Physics Letters A}\ }\textbf {\bibinfo
  {volume} {383}},\ \bibinfo {pages} {3167} (\bibinfo {year}
  {2019})}\BibitemShut {NoStop}%
\bibitem [{\citenamefont {Ceferino}\ \emph {et~al.}(2020)\citenamefont
  {Ceferino}, \citenamefont {Song}, \citenamefont {Magorrian}, \citenamefont
  {Z\'olyomi},\ and\ \citenamefont {Fal'ko}}]{PhysRevB.101.245432}%
  \BibitemOpen
  \bibfield  {author} {\bibinfo {author} {\bibfnamefont {A.}~\bibnamefont
  {Ceferino}}, \bibinfo {author} {\bibfnamefont {K.~W.}\ \bibnamefont {Song}},
  \bibinfo {author} {\bibfnamefont {S.~J.}\ \bibnamefont {Magorrian}}, \bibinfo
  {author} {\bibfnamefont {V.}~\bibnamefont {Z\'olyomi}},\ and\ \bibinfo
  {author} {\bibfnamefont {V.~I.}\ \bibnamefont {Fal'ko}},\ }\bibfield  {title}
  {\bibinfo {title} {Crossover from weakly indirect to direct excitons in
  atomically thin films of inse},\ }\href
  {https://doi.org/10.1103/PhysRevB.101.245432} {\bibfield  {journal} {\bibinfo
   {journal} {Phys. Rev. B}\ }\textbf {\bibinfo {volume} {101}},\ \bibinfo
  {pages} {245432} (\bibinfo {year} {2020})}\BibitemShut {NoStop}%
\bibitem [{\citenamefont {Pierret}\ \emph {et~al.}(2022)\citenamefont
  {Pierret}, \citenamefont {Mele}, \citenamefont {Graef}, \citenamefont
  {Palomo}, \citenamefont {Taniguchi}, \citenamefont {Watanabe}, \citenamefont
  {Li}, \citenamefont {Toury}, \citenamefont {Journet}, \citenamefont {Steyer},
  \citenamefont {Garnier}, \citenamefont {Loiseau}, \citenamefont {Berroir},
  \citenamefont {Bocquillon}, \citenamefont {Fève}, \citenamefont {Voisin},
  \citenamefont {Baudin}, \citenamefont {Rosticher},\ and\ \citenamefont
  {Plaçais}}]{Pierret_2022}%
  \BibitemOpen
  \bibfield  {author} {\bibinfo {author} {\bibfnamefont {A.}~\bibnamefont
  {Pierret}}, \bibinfo {author} {\bibfnamefont {D.}~\bibnamefont {Mele}},
  \bibinfo {author} {\bibfnamefont {H.}~\bibnamefont {Graef}}, \bibinfo
  {author} {\bibfnamefont {J.}~\bibnamefont {Palomo}}, \bibinfo {author}
  {\bibfnamefont {T.}~\bibnamefont {Taniguchi}}, \bibinfo {author}
  {\bibfnamefont {K.}~\bibnamefont {Watanabe}}, \bibinfo {author}
  {\bibfnamefont {Y.}~\bibnamefont {Li}}, \bibinfo {author} {\bibfnamefont
  {B.}~\bibnamefont {Toury}}, \bibinfo {author} {\bibfnamefont
  {C.}~\bibnamefont {Journet}}, \bibinfo {author} {\bibfnamefont
  {P.}~\bibnamefont {Steyer}}, \bibinfo {author} {\bibfnamefont
  {V.}~\bibnamefont {Garnier}}, \bibinfo {author} {\bibfnamefont
  {A.}~\bibnamefont {Loiseau}}, \bibinfo {author} {\bibfnamefont {J.-M.}\
  \bibnamefont {Berroir}}, \bibinfo {author} {\bibfnamefont {E.}~\bibnamefont
  {Bocquillon}}, \bibinfo {author} {\bibfnamefont {G.}~\bibnamefont {Fève}},
  \bibinfo {author} {\bibfnamefont {C.}~\bibnamefont {Voisin}}, \bibinfo
  {author} {\bibfnamefont {E.}~\bibnamefont {Baudin}}, \bibinfo {author}
  {\bibfnamefont {M.}~\bibnamefont {Rosticher}},\ and\ \bibinfo {author}
  {\bibfnamefont {B.}~\bibnamefont {Plaçais}},\ }\bibfield  {title} {\bibinfo
  {title} {Dielectric permittivity, conductivity and breakdown field of
  hexagonal boron nitride},\ }\href {https://doi.org/10.1088/2053-1591/ac4fe1}
  {\bibfield  {journal} {\bibinfo  {journal} {Materials Research Express}\
  }\textbf {\bibinfo {volume} {9}},\ \bibinfo {pages} {065901} (\bibinfo {year}
  {2022})}\BibitemShut {NoStop}%
\bibitem [{\citenamefont {Kyl\"anp\"a\"a}\ and\ \citenamefont
  {Komsa}(2015)}]{PhysRevB.92.205418}%
  \BibitemOpen
  \bibfield  {author} {\bibinfo {author} {\bibfnamefont {I.}~\bibnamefont
  {Kyl\"anp\"a\"a}}\ and\ \bibinfo {author} {\bibfnamefont {H.-P.}\
  \bibnamefont {Komsa}},\ }\bibfield  {title} {\bibinfo {title} {Binding
  energies of exciton complexes in transition metal dichalcogenide monolayers
  and effect of dielectric environment},\ }\href
  {https://doi.org/10.1103/PhysRevB.92.205418} {\bibfield  {journal} {\bibinfo
  {journal} {Phys. Rev. B}\ }\textbf {\bibinfo {volume} {92}},\ \bibinfo
  {pages} {205418} (\bibinfo {year} {2015})}\BibitemShut {NoStop}%
\bibitem [{\citenamefont {Men{\'e}ndez-Proupin}\ \emph
  {et~al.}(2024)\citenamefont {Men{\'e}ndez-Proupin}, \citenamefont {Morell},
  \citenamefont {Marques},\ and\ \citenamefont
  {Trallero-Giner}}]{Menendez-Proupin2024}%
  \BibitemOpen
  \bibfield  {author} {\bibinfo {author} {\bibfnamefont {E.}~\bibnamefont
  {Men{\'e}ndez-Proupin}}, \bibinfo {author} {\bibfnamefont {E.~S.}\
  \bibnamefont {Morell}}, \bibinfo {author} {\bibfnamefont {G.~E.}\
  \bibnamefont {Marques}},\ and\ \bibinfo {author} {\bibfnamefont
  {C.}~\bibnamefont {Trallero-Giner}},\ }\bibfield  {title} {\bibinfo {title}
  {Lattice vibration modes and electron--phonon interactions in monolayer vs.
  bilayer of transition metal dichalcogenides},\ }\href
  {https://doi.org/10.1039/D3RA08759J} {\bibfield  {journal} {\bibinfo
  {journal} {RSC Advances}\ }\textbf {\bibinfo {volume} {14}},\ \bibinfo
  {pages} {5234} (\bibinfo {year} {2024})}\BibitemShut {NoStop}%
\bibitem [{\citenamefont {Das~Sarma}\ and\ \citenamefont
  {Hwang}(2015)}]{DasSarma2015}%
  \BibitemOpen
  \bibfield  {author} {\bibinfo {author} {\bibfnamefont {S.}~\bibnamefont
  {Das~Sarma}}\ and\ \bibinfo {author} {\bibfnamefont {E.~H.}\ \bibnamefont
  {Hwang}},\ }\bibfield  {title} {\bibinfo {title} {Screening and transport in
  2d semiconductor systems at low temperatures},\ }\href
  {https://doi.org/10.1038/srep16655} {\bibfield  {journal} {\bibinfo
  {journal} {Scientific Reports}\ }\textbf {\bibinfo {volume} {5}},\ \bibinfo
  {pages} {16655} (\bibinfo {year} {2015})}\BibitemShut {NoStop}%
\bibitem [{\citenamefont {Chubukov}\ and\ \citenamefont
  {Maslov}(2003)}]{PhysRevB.68.155113}%
  \BibitemOpen
  \bibfield  {author} {\bibinfo {author} {\bibfnamefont {A.~V.}\ \bibnamefont
  {Chubukov}}\ and\ \bibinfo {author} {\bibfnamefont {D.~L.}\ \bibnamefont
  {Maslov}},\ }\bibfield  {title} {\bibinfo {title} {Nonanalytic corrections to
  the fermi-liquid behavior},\ }\href
  {https://doi.org/10.1103/PhysRevB.68.155113} {\bibfield  {journal} {\bibinfo
  {journal} {Phys. Rev. B}\ }\textbf {\bibinfo {volume} {68}},\ \bibinfo
  {pages} {155113} (\bibinfo {year} {2003})}\BibitemShut {NoStop}%
\bibitem [{\citenamefont {Principi}\ \emph {et~al.}(2013)\citenamefont
  {Principi}, \citenamefont {Vignale}, \citenamefont {Carrega},\ and\
  \citenamefont {Polini}}]{PhysRevB.88.195405}%
  \BibitemOpen
  \bibfield  {author} {\bibinfo {author} {\bibfnamefont {A.}~\bibnamefont
  {Principi}}, \bibinfo {author} {\bibfnamefont {G.}~\bibnamefont {Vignale}},
  \bibinfo {author} {\bibfnamefont {M.}~\bibnamefont {Carrega}},\ and\ \bibinfo
  {author} {\bibfnamefont {M.}~\bibnamefont {Polini}},\ }\bibfield  {title}
  {\bibinfo {title} {Intrinsic lifetime of dirac plasmons in graphene},\ }\href
  {https://doi.org/10.1103/PhysRevB.88.195405} {\bibfield  {journal} {\bibinfo
  {journal} {Phys. Rev. B}\ }\textbf {\bibinfo {volume} {88}},\ \bibinfo
  {pages} {195405} (\bibinfo {year} {2013})}\BibitemShut {NoStop}%
\bibitem [{\citenamefont {Chubukov}\ and\ \citenamefont
  {Maslov}(2010)}]{PhysRevB.81.245102}%
  \BibitemOpen
  \bibfield  {author} {\bibinfo {author} {\bibfnamefont {A.~V.}\ \bibnamefont
  {Chubukov}}\ and\ \bibinfo {author} {\bibfnamefont {D.~L.}\ \bibnamefont
  {Maslov}},\ }\bibfield  {title} {\bibinfo {title} {Universal and nonuniversal
  renormalizations in fermi liquids},\ }\href
  {https://doi.org/10.1103/PhysRevB.81.245102} {\bibfield  {journal} {\bibinfo
  {journal} {Phys. Rev. B}\ }\textbf {\bibinfo {volume} {81}},\ \bibinfo
  {pages} {245102} (\bibinfo {year} {2010})}\BibitemShut {NoStop}%
\bibitem [{\citenamefont {Gindikin}\ and\ \citenamefont
  {Chubukov}(2024)}]{PhysRevB.109.115156}%
  \BibitemOpen
  \bibfield  {author} {\bibinfo {author} {\bibfnamefont {Y.}~\bibnamefont
  {Gindikin}}\ and\ \bibinfo {author} {\bibfnamefont {A.~V.}\ \bibnamefont
  {Chubukov}},\ }\bibfield  {title} {\bibinfo {title} {Fermi surface geometry
  and optical conductivity of a two-dimensional electron gas near an
  ising-nematic quantum critical point},\ }\href
  {https://doi.org/10.1103/PhysRevB.109.115156} {\bibfield  {journal} {\bibinfo
   {journal} {Phys. Rev. B}\ }\textbf {\bibinfo {volume} {109}},\ \bibinfo
  {pages} {115156} (\bibinfo {year} {2024})}\BibitemShut {NoStop}%
\bibitem [{\citenamefont {Ando}\ \emph {et~al.}(1982)\citenamefont {Ando},
  \citenamefont {Fowler},\ and\ \citenamefont {Stern}}]{RevModPhys.54.437}%
  \BibitemOpen
  \bibfield  {author} {\bibinfo {author} {\bibfnamefont {T.}~\bibnamefont
  {Ando}}, \bibinfo {author} {\bibfnamefont {A.~B.}\ \bibnamefont {Fowler}},\
  and\ \bibinfo {author} {\bibfnamefont {F.}~\bibnamefont {Stern}},\ }\bibfield
   {title} {\bibinfo {title} {Electronic properties of two-dimensional
  systems},\ }\href {https://doi.org/10.1103/RevModPhys.54.437} {\bibfield
  {journal} {\bibinfo  {journal} {Rev. Mod. Phys.}\ }\textbf {\bibinfo {volume}
  {54}},\ \bibinfo {pages} {437} (\bibinfo {year} {1982})}\BibitemShut
  {NoStop}%
\bibitem [{\citenamefont {Azadi}\ \emph {et~al.}(2025)\citenamefont {Azadi},
  \citenamefont {Drummond}, \citenamefont {Principi}, \citenamefont
  {Belosludov},\ and\ \citenamefont {Bahramy}}]{3wjg-y6qk}%
  \BibitemOpen
  \bibfield  {author} {\bibinfo {author} {\bibfnamefont {S.}~\bibnamefont
  {Azadi}}, \bibinfo {author} {\bibfnamefont {N.~D.}\ \bibnamefont {Drummond}},
  \bibinfo {author} {\bibfnamefont {A.}~\bibnamefont {Principi}}, \bibinfo
  {author} {\bibfnamefont {R.~V.}\ \bibnamefont {Belosludov}},\ and\ \bibinfo
  {author} {\bibfnamefont {M.~S.}\ \bibnamefont {Bahramy}},\ }\bibfield
  {title} {\bibinfo {title} {Quantum monte carlo study of the quasiparticle
  effective mass of the two-dimensional uniform electron liquid},\ }\href
  {https://doi.org/10.1103/3wjg-y6qk} {\bibfield  {journal} {\bibinfo
  {journal} {Phys. Rev. B}\ }\textbf {\bibinfo {volume} {112}},\ \bibinfo
  {pages} {075141} (\bibinfo {year} {2025})}\BibitemShut {NoStop}%
\bibitem [{\citenamefont {Echeverry}\ \emph {et~al.}(2016)\citenamefont
  {Echeverry}, \citenamefont {Urbaszek}, \citenamefont {Amand}, \citenamefont
  {Marie},\ and\ \citenamefont {Gerber}}]{PhysRevB.93.121107}%
  \BibitemOpen
  \bibfield  {author} {\bibinfo {author} {\bibfnamefont {J.~P.}\ \bibnamefont
  {Echeverry}}, \bibinfo {author} {\bibfnamefont {B.}~\bibnamefont {Urbaszek}},
  \bibinfo {author} {\bibfnamefont {T.}~\bibnamefont {Amand}}, \bibinfo
  {author} {\bibfnamefont {X.}~\bibnamefont {Marie}},\ and\ \bibinfo {author}
  {\bibfnamefont {I.~C.}\ \bibnamefont {Gerber}},\ }\bibfield  {title}
  {\bibinfo {title} {Splitting between bright and dark excitons in transition
  metal dichalcogenide monolayers},\ }\href
  {https://doi.org/10.1103/PhysRevB.93.121107} {\bibfield  {journal} {\bibinfo
  {journal} {Phys. Rev. B}\ }\textbf {\bibinfo {volume} {93}},\ \bibinfo
  {pages} {121107} (\bibinfo {year} {2016})}\BibitemShut {NoStop}%
\bibitem [{\citenamefont {Gjerding}\ \emph {et~al.}(2021)\citenamefont
  {Gjerding}, \citenamefont {Taghizadeh}, \citenamefont {Rasmussen},
  \citenamefont {Ali}, \citenamefont {Bertoldo}, \citenamefont {Deilmann},
  \citenamefont {Knøsgaard}, \citenamefont {Kruse}, \citenamefont {Larsen},
  \citenamefont {Manti}, \citenamefont {Pedersen}, \citenamefont {Petralanda},
  \citenamefont {Skovhus}, \citenamefont {Svendsen}, \citenamefont {Mortensen},
  \citenamefont {Olsen},\ and\ \citenamefont {Thygesen}}]{Gjerding_2021}%
  \BibitemOpen
  \bibfield  {author} {\bibinfo {author} {\bibfnamefont {M.~N.}\ \bibnamefont
  {Gjerding}}, \bibinfo {author} {\bibfnamefont {A.}~\bibnamefont
  {Taghizadeh}}, \bibinfo {author} {\bibfnamefont {A.}~\bibnamefont
  {Rasmussen}}, \bibinfo {author} {\bibfnamefont {S.}~\bibnamefont {Ali}},
  \bibinfo {author} {\bibfnamefont {F.}~\bibnamefont {Bertoldo}}, \bibinfo
  {author} {\bibfnamefont {T.}~\bibnamefont {Deilmann}}, \bibinfo {author}
  {\bibfnamefont {N.~R.}\ \bibnamefont {Knøsgaard}}, \bibinfo {author}
  {\bibfnamefont {M.}~\bibnamefont {Kruse}}, \bibinfo {author} {\bibfnamefont
  {A.~H.}\ \bibnamefont {Larsen}}, \bibinfo {author} {\bibfnamefont
  {S.}~\bibnamefont {Manti}}, \bibinfo {author} {\bibfnamefont {T.~G.}\
  \bibnamefont {Pedersen}}, \bibinfo {author} {\bibfnamefont {U.}~\bibnamefont
  {Petralanda}}, \bibinfo {author} {\bibfnamefont {T.}~\bibnamefont {Skovhus}},
  \bibinfo {author} {\bibfnamefont {M.~K.}\ \bibnamefont {Svendsen}}, \bibinfo
  {author} {\bibfnamefont {J.~J.}\ \bibnamefont {Mortensen}}, \bibinfo {author}
  {\bibfnamefont {T.}~\bibnamefont {Olsen}},\ and\ \bibinfo {author}
  {\bibfnamefont {K.~S.}\ \bibnamefont {Thygesen}},\ }\bibfield  {title}
  {\bibinfo {title} {Recent progress of the computational 2d materials database
  (c2db)},\ }\href {https://doi.org/10.1088/2053-1583/ac1059} {\bibfield
  {journal} {\bibinfo  {journal} {2D Materials}\ }\textbf {\bibinfo {volume}
  {8}},\ \bibinfo {pages} {044002} (\bibinfo {year} {2021})}\BibitemShut
  {NoStop}%
\bibitem [{Note1()}]{Note1}%
  \BibitemOpen
  \bibinfo {note} {This value can be compared with optical data on
  hBN-encapsulated MoS$_2$ monolayers~\cite {Robert2020}. Unlike
  magnetotransport, where many-body interactions renormalize effective masses
  and SOS, optical spectroscopy probes excitons without free carriers and thus
  reflects nearly unrenormalized single-particle parameters. Using the
  framework of Ref.~\cite {PhysRevB.95.081301} and the code of Ref.~\cite
  {PhysRevB.96.075431}, we recalculate the bare $\Delta _0$ needed to reproduce
  the reported bright–dark exciton splitting $\Delta _{BD}=14$~meV~\cite
  {Robert2020}, which depends on \begin {equation} \Delta _{BD} = -\Delta _0 +
  \Delta _{\protect \rm bind} + \Delta _{\protect \rm exch}, \end {equation}
  with $\Delta _{\protect \rm bind}$ determined by the effective masses $m_l$,
  $m_u$, and $m_h$, and $\Delta _{\protect \rm exch}$ the short-range exciton
  exchange. Taking $m_h \approx 0.6 m_e$~\cite {Eknapakul2014}, we obtain
  $\Delta _{\protect \rm bind} \approx 19.5$~meV and $\Delta _{\protect \rm
  exch} \approx 2.6$~meV.}\BibitemShut {Stop}%
\bibitem [{\citenamefont {Liu}\ \emph {et~al.}(2015)\citenamefont {Liu},
  \citenamefont {Xiao}, \citenamefont {Yao}, \citenamefont {Xu},\ and\
  \citenamefont {Yao}}]{C4CS00301B}%
  \BibitemOpen
  \bibfield  {author} {\bibinfo {author} {\bibfnamefont {G.-B.}\ \bibnamefont
  {Liu}}, \bibinfo {author} {\bibfnamefont {D.}~\bibnamefont {Xiao}}, \bibinfo
  {author} {\bibfnamefont {Y.}~\bibnamefont {Yao}}, \bibinfo {author}
  {\bibfnamefont {X.}~\bibnamefont {Xu}},\ and\ \bibinfo {author}
  {\bibfnamefont {W.}~\bibnamefont {Yao}},\ }\bibfield  {title} {\bibinfo
  {title} {Electronic structures and theoretical modelling of two-dimensional
  group-vib transition metal dichalcogenides},\ }\href
  {https://doi.org/10.1039/C4CS00301B} {\bibfield  {journal} {\bibinfo
  {journal} {Chem. Soc. Rev.}\ }\textbf {\bibinfo {volume} {44}},\ \bibinfo
  {pages} {2643} (\bibinfo {year} {2015})}\BibitemShut {NoStop}%
\bibitem [{\citenamefont {Pisoni}\ \emph {et~al.}(2019)\citenamefont {Pisoni},
  \citenamefont {Davatz}, \citenamefont {Watanabe}, \citenamefont {Taniguchi},
  \citenamefont {Ihn},\ and\ \citenamefont {Ensslin}}]{PhysRevLett.123.117702}%
  \BibitemOpen
  \bibfield  {author} {\bibinfo {author} {\bibfnamefont {R.}~\bibnamefont
  {Pisoni}}, \bibinfo {author} {\bibfnamefont {T.}~\bibnamefont {Davatz}},
  \bibinfo {author} {\bibfnamefont {K.}~\bibnamefont {Watanabe}}, \bibinfo
  {author} {\bibfnamefont {T.}~\bibnamefont {Taniguchi}}, \bibinfo {author}
  {\bibfnamefont {T.}~\bibnamefont {Ihn}},\ and\ \bibinfo {author}
  {\bibfnamefont {K.}~\bibnamefont {Ensslin}},\ }\bibfield  {title} {\bibinfo
  {title} {Absence of interlayer tunnel coupling of $k$-valley electrons in
  bilayer ${\mathrm{mos}}_{2}$},\ }\href
  {https://doi.org/10.1103/PhysRevLett.123.117702} {\bibfield  {journal}
  {\bibinfo  {journal} {Phys. Rev. Lett.}\ }\textbf {\bibinfo {volume} {123}},\
  \bibinfo {pages} {117702} (\bibinfo {year} {2019})}\BibitemShut {NoStop}%
\bibitem [{\citenamefont {Masseroni}\ \emph {et~al.}(2023)\citenamefont
  {Masseroni}, \citenamefont {Qu}, \citenamefont {Taniguchi}, \citenamefont
  {Watanabe}, \citenamefont {Ihn},\ and\ \citenamefont
  {Ensslin}}]{PhysRevResearch.5.013113}%
  \BibitemOpen
  \bibfield  {author} {\bibinfo {author} {\bibfnamefont {M.}~\bibnamefont
  {Masseroni}}, \bibinfo {author} {\bibfnamefont {T.}~\bibnamefont {Qu}},
  \bibinfo {author} {\bibfnamefont {T.}~\bibnamefont {Taniguchi}}, \bibinfo
  {author} {\bibfnamefont {K.}~\bibnamefont {Watanabe}}, \bibinfo {author}
  {\bibfnamefont {T.}~\bibnamefont {Ihn}},\ and\ \bibinfo {author}
  {\bibfnamefont {K.}~\bibnamefont {Ensslin}},\ }\bibfield  {title} {\bibinfo
  {title} {Evidence of the coulomb gap in the density of states of
  ${\mathrm{mos}}_{2}$},\ }\href
  {https://doi.org/10.1103/PhysRevResearch.5.013113} {\bibfield  {journal}
  {\bibinfo  {journal} {Phys. Rev. Res.}\ }\textbf {\bibinfo {volume} {5}},\
  \bibinfo {pages} {013113} (\bibinfo {year} {2023})}\BibitemShut {NoStop}%
\bibitem [{\citenamefont {Masseroni}\ \emph {et~al.}(2021)\citenamefont
  {Masseroni}, \citenamefont {Davatz}, \citenamefont {Pisoni}, \citenamefont
  {de~Vries}, \citenamefont {Rickhaus}, \citenamefont {Taniguchi},
  \citenamefont {Watanabe}, \citenamefont {Fal'ko}, \citenamefont {Ihn},\ and\
  \citenamefont {Ensslin}}]{PhysRevResearch.3.023047}%
  \BibitemOpen
  \bibfield  {author} {\bibinfo {author} {\bibfnamefont {M.}~\bibnamefont
  {Masseroni}}, \bibinfo {author} {\bibfnamefont {T.}~\bibnamefont {Davatz}},
  \bibinfo {author} {\bibfnamefont {R.}~\bibnamefont {Pisoni}}, \bibinfo
  {author} {\bibfnamefont {F.~K.}\ \bibnamefont {de~Vries}}, \bibinfo {author}
  {\bibfnamefont {P.}~\bibnamefont {Rickhaus}}, \bibinfo {author}
  {\bibfnamefont {T.}~\bibnamefont {Taniguchi}}, \bibinfo {author}
  {\bibfnamefont {K.}~\bibnamefont {Watanabe}}, \bibinfo {author}
  {\bibfnamefont {V.}~\bibnamefont {Fal'ko}}, \bibinfo {author} {\bibfnamefont
  {T.}~\bibnamefont {Ihn}},\ and\ \bibinfo {author} {\bibfnamefont
  {K.}~\bibnamefont {Ensslin}},\ }\bibfield  {title} {\bibinfo {title}
  {Electron transport in dual-gated three-layer
  $\mathrm{Mo}{\mathrm{s}}_{2}$},\ }\href
  {https://doi.org/10.1103/PhysRevResearch.3.023047} {\bibfield  {journal}
  {\bibinfo  {journal} {Phys. Rev. Res.}\ }\textbf {\bibinfo {volume} {3}},\
  \bibinfo {pages} {023047} (\bibinfo {year} {2021})}\BibitemShut {NoStop}%
\bibitem [{\citenamefont {Masseroni}(2024)}]{Masseroni2024}%
  \BibitemOpen
  \bibfield  {author} {\bibinfo {author} {\bibfnamefont {M.}~\bibnamefont
  {Masseroni}},\ }\emph {\bibinfo {title} {Electronic transport experiments in
  2D materials with spin-orbit coupling}},\ \href
  {https://doi.org/10.3929/ethz-b-000701916} {Ph.D. thesis},\ \bibinfo
  {school} {ETH Zurich} (\bibinfo {year} {2024})\BibitemShut {NoStop}%
\bibitem [{\citenamefont {Masseroni}\ \emph {et~al.}(2025)\citenamefont
  {Masseroni}, \citenamefont {Soltero}, \citenamefont {McHugh}, \citenamefont
  {Rozhansky}, \citenamefont {Li}, \citenamefont {Schmidhuber}, \citenamefont
  {Niese}, \citenamefont {Taniguchi}, \citenamefont {Watanabe}, \citenamefont
  {Fal'ko}, \citenamefont {Ihn},\ and\ \citenamefont
  {Ensslin}}]{Masseroni2025}%
  \BibitemOpen
  \bibfield  {author} {\bibinfo {author} {\bibfnamefont {M.}~\bibnamefont
  {Masseroni}}, \bibinfo {author} {\bibfnamefont {I.}~\bibnamefont {Soltero}},
  \bibinfo {author} {\bibfnamefont {J.~G.}\ \bibnamefont {McHugh}}, \bibinfo
  {author} {\bibfnamefont {I.}~\bibnamefont {Rozhansky}}, \bibinfo {author}
  {\bibfnamefont {X.}~\bibnamefont {Li}}, \bibinfo {author} {\bibfnamefont
  {A.}~\bibnamefont {Schmidhuber}}, \bibinfo {author} {\bibfnamefont
  {M.}~\bibnamefont {Niese}}, \bibinfo {author} {\bibfnamefont
  {T.}~\bibnamefont {Taniguchi}}, \bibinfo {author} {\bibfnamefont
  {K.}~\bibnamefont {Watanabe}}, \bibinfo {author} {\bibfnamefont {V.~I.}\
  \bibnamefont {Fal'ko}}, \bibinfo {author} {\bibfnamefont {T.}~\bibnamefont
  {Ihn}},\ and\ \bibinfo {author} {\bibfnamefont {K.}~\bibnamefont {Ensslin}},\
  }\bibfield  {title} {\bibinfo {title} {Gate-tunable band edge in few-layer
  mos2},\ }\href {https://doi.org/10.1021/acs.nanolett.5c01998} {\bibfield
  {journal} {\bibinfo  {journal} {Nano Letters}\ }\textbf {\bibinfo {volume}
  {25}},\ \bibinfo {pages} {10472} (\bibinfo {year} {2025})}\BibitemShut
  {NoStop}%
\bibitem [{\citenamefont {Campo}\ and\ \citenamefont
  {Cococcioni}(2010)}]{campo2010extended}%
  \BibitemOpen
  \bibfield  {author} {\bibinfo {author} {\bibfnamefont {V.~L.}\ \bibnamefont
  {Campo}}\ and\ \bibinfo {author} {\bibfnamefont {M.}~\bibnamefont
  {Cococcioni}},\ }\bibfield  {title} {\bibinfo {title} {Extended dft+ u+ v
  method with on-site and inter-site electronic interactions},\ }\href@noop {}
  {\bibfield  {journal} {\bibinfo  {journal} {Journal of Physics: Condensed
  Matter}\ }\textbf {\bibinfo {volume} {22}},\ \bibinfo {pages} {055602}
  (\bibinfo {year} {2010})}\BibitemShut {NoStop}%
\bibitem [{\citenamefont {Mak}\ \emph {et~al.}(2012)\citenamefont {Mak},
  \citenamefont {He}, \citenamefont {Shan},\ and\ \citenamefont
  {Heinz}}]{Mak2012}%
  \BibitemOpen
  \bibfield  {author} {\bibinfo {author} {\bibfnamefont {K.~F.}\ \bibnamefont
  {Mak}}, \bibinfo {author} {\bibfnamefont {K.}~\bibnamefont {He}}, \bibinfo
  {author} {\bibfnamefont {J.}~\bibnamefont {Shan}},\ and\ \bibinfo {author}
  {\bibfnamefont {T.~F.}\ \bibnamefont {Heinz}},\ }\bibfield  {title} {\bibinfo
  {title} {{Control of valley polarization in monolayer MoS2 by optical
  helicity}},\ }\href {https://doi.org/10.1038/nnano.2012.96} {\bibfield
  {journal} {\bibinfo  {journal} {Nature Nanotechnology}\ }\textbf {\bibinfo
  {volume} {7}},\ \bibinfo {pages} {494} (\bibinfo {year} {2012})}\BibitemShut
  {NoStop}%
\bibitem [{\citenamefont {Klots}\ \emph {et~al.}(2014)\citenamefont {Klots},
  \citenamefont {Newaz}, \citenamefont {Wang}, \citenamefont {Prasai},
  \citenamefont {Krzyzanowska}, \citenamefont {Lin}, \citenamefont {Caudel},
  \citenamefont {Ghimire}, \citenamefont {Yan}, \citenamefont {Ivanov},
  \citenamefont {Velizhanin}, \citenamefont {Burger}, \citenamefont {Mandrus},
  \citenamefont {Tolk}, \citenamefont {Pantelides},\ and\ \citenamefont
  {Bolotin}}]{Klots2014}%
  \BibitemOpen
  \bibfield  {author} {\bibinfo {author} {\bibfnamefont {A.~R.}\ \bibnamefont
  {Klots}}, \bibinfo {author} {\bibfnamefont {A.~K.~M.}\ \bibnamefont {Newaz}},
  \bibinfo {author} {\bibfnamefont {B.}~\bibnamefont {Wang}}, \bibinfo {author}
  {\bibfnamefont {D.}~\bibnamefont {Prasai}}, \bibinfo {author} {\bibfnamefont
  {H.}~\bibnamefont {Krzyzanowska}}, \bibinfo {author} {\bibfnamefont
  {J.}~\bibnamefont {Lin}}, \bibinfo {author} {\bibfnamefont {D.}~\bibnamefont
  {Caudel}}, \bibinfo {author} {\bibfnamefont {N.~J.}\ \bibnamefont {Ghimire}},
  \bibinfo {author} {\bibfnamefont {J.}~\bibnamefont {Yan}}, \bibinfo {author}
  {\bibfnamefont {B.~L.}\ \bibnamefont {Ivanov}}, \bibinfo {author}
  {\bibfnamefont {K.~A.}\ \bibnamefont {Velizhanin}}, \bibinfo {author}
  {\bibfnamefont {A.}~\bibnamefont {Burger}}, \bibinfo {author} {\bibfnamefont
  {D.~G.}\ \bibnamefont {Mandrus}}, \bibinfo {author} {\bibfnamefont {N.~H.}\
  \bibnamefont {Tolk}}, \bibinfo {author} {\bibfnamefont {S.~T.}\ \bibnamefont
  {Pantelides}},\ and\ \bibinfo {author} {\bibfnamefont {K.~I.}\ \bibnamefont
  {Bolotin}},\ }\bibfield  {title} {\bibinfo {title} {Probing excitonic states
  in suspended two-dimensional semiconductors by photocurrent spectroscopy},\
  }\href {https://doi.org/10.1038/srep06608} {\bibfield  {journal} {\bibinfo
  {journal} {Scientific Reports}\ }\textbf {\bibinfo {volume} {4}},\ \bibinfo
  {pages} {6608} (\bibinfo {year} {2014})}\BibitemShut {NoStop}%
\bibitem [{\citenamefont {Li}\ \emph {et~al.}(2014)\citenamefont {Li},
  \citenamefont {Birdwell}, \citenamefont {Amani}, \citenamefont {Burke},
  \citenamefont {Ling}, \citenamefont {Lee}, \citenamefont {Liang},
  \citenamefont {Peng}, \citenamefont {Richter}, \citenamefont {Kong},
  \citenamefont {Gundlach},\ and\ \citenamefont {Nguyen}}]{PhysRevB.90.195434}%
  \BibitemOpen
  \bibfield  {author} {\bibinfo {author} {\bibfnamefont {W.}~\bibnamefont
  {Li}}, \bibinfo {author} {\bibfnamefont {A.~G.}\ \bibnamefont {Birdwell}},
  \bibinfo {author} {\bibfnamefont {M.}~\bibnamefont {Amani}}, \bibinfo
  {author} {\bibfnamefont {R.~A.}\ \bibnamefont {Burke}}, \bibinfo {author}
  {\bibfnamefont {X.}~\bibnamefont {Ling}}, \bibinfo {author} {\bibfnamefont
  {Y.-H.}\ \bibnamefont {Lee}}, \bibinfo {author} {\bibfnamefont
  {X.}~\bibnamefont {Liang}}, \bibinfo {author} {\bibfnamefont
  {L.}~\bibnamefont {Peng}}, \bibinfo {author} {\bibfnamefont {C.~A.}\
  \bibnamefont {Richter}}, \bibinfo {author} {\bibfnamefont {J.}~\bibnamefont
  {Kong}}, \bibinfo {author} {\bibfnamefont {D.~J.}\ \bibnamefont {Gundlach}},\
  and\ \bibinfo {author} {\bibfnamefont {N.~V.}\ \bibnamefont {Nguyen}},\
  }\bibfield  {title} {\bibinfo {title} {Broadband optical properties of
  large-area monolayer cvd molybdenum disulfide},\ }\href
  {https://doi.org/10.1103/PhysRevB.90.195434} {\bibfield  {journal} {\bibinfo
  {journal} {Phys. Rev. B}\ }\textbf {\bibinfo {volume} {90}},\ \bibinfo
  {pages} {195434} (\bibinfo {year} {2014})}\BibitemShut {NoStop}%
\bibitem [{\citenamefont {Kozawa}\ \emph {et~al.}(2014)\citenamefont {Kozawa},
  \citenamefont {Kumar}, \citenamefont {Carvalho}, \citenamefont {Kumar~Amara},
  \citenamefont {Zhao}, \citenamefont {Wang}, \citenamefont {Toh},
  \citenamefont {Ribeiro}, \citenamefont {Castro~Neto}, \citenamefont
  {Matsuda},\ and\ \citenamefont {Eda}}]{Kozawa2014}%
  \BibitemOpen
  \bibfield  {author} {\bibinfo {author} {\bibfnamefont {D.}~\bibnamefont
  {Kozawa}}, \bibinfo {author} {\bibfnamefont {R.}~\bibnamefont {Kumar}},
  \bibinfo {author} {\bibfnamefont {A.}~\bibnamefont {Carvalho}}, \bibinfo
  {author} {\bibfnamefont {K.}~\bibnamefont {Kumar~Amara}}, \bibinfo {author}
  {\bibfnamefont {W.}~\bibnamefont {Zhao}}, \bibinfo {author} {\bibfnamefont
  {S.}~\bibnamefont {Wang}}, \bibinfo {author} {\bibfnamefont {M.}~\bibnamefont
  {Toh}}, \bibinfo {author} {\bibfnamefont {R.~M.}\ \bibnamefont {Ribeiro}},
  \bibinfo {author} {\bibfnamefont {A.~H.}\ \bibnamefont {Castro~Neto}},
  \bibinfo {author} {\bibfnamefont {K.}~\bibnamefont {Matsuda}},\ and\ \bibinfo
  {author} {\bibfnamefont {G.}~\bibnamefont {Eda}},\ }\bibfield  {title}
  {\bibinfo {title} {Photocarrier relaxation pathway in two-dimensional
  semiconducting transition metal dichalcogenides},\ }\href
  {https://doi.org/10.1038/ncomms5543} {\bibfield  {journal} {\bibinfo
  {journal} {Nature Communications}\ }\textbf {\bibinfo {volume} {5}},\
  \bibinfo {pages} {4543} (\bibinfo {year} {2014})}\BibitemShut {NoStop}%
\bibitem [{\citenamefont {Perdew}\ and\ \citenamefont
  {Levy}(1983)}]{PhysRevLett.51.1884}%
  \BibitemOpen
  \bibfield  {author} {\bibinfo {author} {\bibfnamefont {J.~P.}\ \bibnamefont
  {Perdew}}\ and\ \bibinfo {author} {\bibfnamefont {M.}~\bibnamefont {Levy}},\
  }\bibfield  {title} {\bibinfo {title} {Physical content of the exact
  kohn-sham orbital energies: Band gaps and derivative discontinuities},\
  }\href {https://doi.org/10.1103/PhysRevLett.51.1884} {\bibfield  {journal}
  {\bibinfo  {journal} {Phys. Rev. Lett.}\ }\textbf {\bibinfo {volume} {51}},\
  \bibinfo {pages} {1884} (\bibinfo {year} {1983})}\BibitemShut {NoStop}%
\bibitem [{\citenamefont {Sampanraj}\ \emph {et~al.}(2025)\citenamefont
  {Sampanraj}, \citenamefont {Devendar}, \citenamefont {Pothal}, \citenamefont
  {Jaiswal},\ and\ \citenamefont {Balakrishnan}}]{PhysRevB.111.075423}%
  \BibitemOpen
  \bibfield  {author} {\bibinfo {author} {\bibfnamefont {A.}~\bibnamefont
  {Sampanraj}}, \bibinfo {author} {\bibfnamefont {L.}~\bibnamefont {Devendar}},
  \bibinfo {author} {\bibfnamefont {B.}~\bibnamefont {Pothal}}, \bibinfo
  {author} {\bibfnamefont {M.}~\bibnamefont {Jaiswal}},\ and\ \bibinfo {author}
  {\bibfnamefont {J.}~\bibnamefont {Balakrishnan}},\ }\bibfield  {title}
  {\bibinfo {title} {Interface-controlled thermal transport in monolayer
  molybdenum disulfide},\ }\href {https://doi.org/10.1103/PhysRevB.111.075423}
  {\bibfield  {journal} {\bibinfo  {journal} {Phys. Rev. B}\ }\textbf {\bibinfo
  {volume} {111}},\ \bibinfo {pages} {075423} (\bibinfo {year}
  {2025})}\BibitemShut {NoStop}%
\bibitem [{\citenamefont {Mak}\ \emph {et~al.}(2010)\citenamefont {Mak},
  \citenamefont {Lee}, \citenamefont {Hone}, \citenamefont {Shan},\ and\
  \citenamefont {Heinz}}]{PhysRevLett.105.136805}%
  \BibitemOpen
  \bibfield  {author} {\bibinfo {author} {\bibfnamefont {K.~F.}\ \bibnamefont
  {Mak}}, \bibinfo {author} {\bibfnamefont {C.}~\bibnamefont {Lee}}, \bibinfo
  {author} {\bibfnamefont {J.}~\bibnamefont {Hone}}, \bibinfo {author}
  {\bibfnamefont {J.}~\bibnamefont {Shan}},\ and\ \bibinfo {author}
  {\bibfnamefont {T.~F.}\ \bibnamefont {Heinz}},\ }\bibfield  {title} {\bibinfo
  {title} {Atomically thin ${\mathrm{mos}}_{2}$: A new direct-gap
  semiconductor},\ }\href {https://doi.org/10.1103/PhysRevLett.105.136805}
  {\bibfield  {journal} {\bibinfo  {journal} {Phys. Rev. Lett.}\ }\textbf
  {\bibinfo {volume} {105}},\ \bibinfo {pages} {136805} (\bibinfo {year}
  {2010})}\BibitemShut {NoStop}%
\bibitem [{\citenamefont {Ramezani}\ \emph {et~al.}(2024)\citenamefont
  {Ramezani}, \citenamefont {\ifmmode \mbox{\c{S}}\else \c{S}\fi{}a\ifmmode
  \mbox{\c{s}}\else \c{s}\fi{}\ifmmode \imath \else \i
  \fi{}o\ifmmode~\breve{g}\else \u{g}\fi{}lu}, \citenamefont {Hadipour},
  \citenamefont {Soleimani}, \citenamefont {Friedrich}, \citenamefont
  {Bl\"ugel},\ and\ \citenamefont {Mertig}}]{PhysRevB.109.125108}%
  \BibitemOpen
  \bibfield  {author} {\bibinfo {author} {\bibfnamefont {H.~R.}\ \bibnamefont
  {Ramezani}}, \bibinfo {author} {\bibfnamefont {E.}~\bibnamefont {\ifmmode
  \mbox{\c{S}}\else \c{S}\fi{}a\ifmmode \mbox{\c{s}}\else \c{s}\fi{}\ifmmode
  \imath \else \i \fi{}o\ifmmode~\breve{g}\else \u{g}\fi{}lu}}, \bibinfo
  {author} {\bibfnamefont {H.}~\bibnamefont {Hadipour}}, \bibinfo {author}
  {\bibfnamefont {H.~R.}\ \bibnamefont {Soleimani}}, \bibinfo {author}
  {\bibfnamefont {C.}~\bibnamefont {Friedrich}}, \bibinfo {author}
  {\bibfnamefont {S.}~\bibnamefont {Bl\"ugel}},\ and\ \bibinfo {author}
  {\bibfnamefont {I.}~\bibnamefont {Mertig}},\ }\bibfield  {title} {\bibinfo
  {title} {Nonconventional screening of coulomb interaction in two-dimensional
  semiconductors and metals: A comprehensive constrained random phase
  approximation study of $m{X}_{2}$ $(m=\mathrm{Mo}, \mathrm{W}, \mathrm{Nb},
  \mathrm{Ta}; x\mathrm{=}\mathrm{S}, \mathrm{Se}, \mathrm{Te})$},\ }\href
  {https://doi.org/10.1103/PhysRevB.109.125108} {\bibfield  {journal} {\bibinfo
   {journal} {Phys. Rev. B}\ }\textbf {\bibinfo {volume} {109}},\ \bibinfo
  {pages} {125108} (\bibinfo {year} {2024})}\BibitemShut {NoStop}%
\bibitem [{\citenamefont {Szyniszewski}\ \emph {et~al.}(2017)\citenamefont
  {Szyniszewski}, \citenamefont {Mostaani}, \citenamefont {Drummond},\ and\
  \citenamefont {Fal'ko}}]{PhysRevB.95.081301}%
  \BibitemOpen
  \bibfield  {author} {\bibinfo {author} {\bibfnamefont {M.}~\bibnamefont
  {Szyniszewski}}, \bibinfo {author} {\bibfnamefont {E.}~\bibnamefont
  {Mostaani}}, \bibinfo {author} {\bibfnamefont {N.~D.}\ \bibnamefont
  {Drummond}},\ and\ \bibinfo {author} {\bibfnamefont {V.~I.}\ \bibnamefont
  {Fal'ko}},\ }\bibfield  {title} {\bibinfo {title} {Binding energies of trions
  and biexcitons in two-dimensional semiconductors from diffusion quantum monte
  carlo calculations},\ }\href {https://doi.org/10.1103/PhysRevB.95.081301}
  {\bibfield  {journal} {\bibinfo  {journal} {Phys. Rev. B}\ }\textbf {\bibinfo
  {volume} {95}},\ \bibinfo {pages} {081301} (\bibinfo {year}
  {2017})}\BibitemShut {NoStop}%
\bibitem [{\citenamefont {Mostaani}\ \emph {et~al.}(2017)\citenamefont
  {Mostaani}, \citenamefont {Szyniszewski}, \citenamefont {Price},
  \citenamefont {Maezono}, \citenamefont {Danovich}, \citenamefont {Hunt},
  \citenamefont {Drummond},\ and\ \citenamefont {Fal'ko}}]{PhysRevB.96.075431}%
  \BibitemOpen
  \bibfield  {author} {\bibinfo {author} {\bibfnamefont {E.}~\bibnamefont
  {Mostaani}}, \bibinfo {author} {\bibfnamefont {M.}~\bibnamefont
  {Szyniszewski}}, \bibinfo {author} {\bibfnamefont {C.~H.}\ \bibnamefont
  {Price}}, \bibinfo {author} {\bibfnamefont {R.}~\bibnamefont {Maezono}},
  \bibinfo {author} {\bibfnamefont {M.}~\bibnamefont {Danovich}}, \bibinfo
  {author} {\bibfnamefont {R.~J.}\ \bibnamefont {Hunt}}, \bibinfo {author}
  {\bibfnamefont {N.~D.}\ \bibnamefont {Drummond}},\ and\ \bibinfo {author}
  {\bibfnamefont {V.~I.}\ \bibnamefont {Fal'ko}},\ }\bibfield  {title}
  {\bibinfo {title} {Diffusion quantum monte carlo study of excitonic complexes
  in two-dimensional transition-metal dichalcogenides},\ }\href
  {https://doi.org/10.1103/PhysRevB.96.075431} {\bibfield  {journal} {\bibinfo
  {journal} {Phys. Rev. B}\ }\textbf {\bibinfo {volume} {96}},\ \bibinfo
  {pages} {075431} (\bibinfo {year} {2017})}\BibitemShut {NoStop}%
\end{thebibliography}%

\end{document}